\documentclass[conference]{IEEEtran}
\usepackage{cite}
\usepackage{amsmath,amssymb,amsfonts}
\usepackage{algorithmic}
\usepackage{graphicx}
\usepackage{textcomp}
\usepackage{xcolor}
\usepackage[hyphens]{url}

\def\BibTeX{{\rm B\kern-.05em{\sc i\kern-.025em b}\kern-.08em
    T\kern-.1667em\lower.7ex\hbox{E}\kern-.125emX}}

\usepackage{color}
\usepackage{soul}
\usepackage{multirow}
\usepackage{makecell, rotating}
\usepackage{subfigure}
\usepackage{enumerate}
\usepackage{enumitem}
\usepackage{balance}
\usepackage[linesnumbered,ruled,lined]{algorithm2e}
\usepackage{threeparttable}
\newcommand{\RebuttalChange}[1]{\textcolor{black}{#1}}

\newcommand{\SolutionName}{TensorFHE}
\newcommand{\Tofill}[1]{\textcolor{black}{#1}}

\IEEEoverridecommandlockouts

\title{\SolutionName: Achieving Practical Computation\\on Encrypted Data Using GPGPU} 
\author{
    
        Shengyu Fan\IEEEauthorrefmark{1}\IEEEauthorrefmark{3}, 
        Zhiwei Wang\IEEEauthorrefmark{1}\IEEEauthorrefmark{2}, 
        Weizhi Xu\IEEEauthorrefmark{3}, 
        Rui Hou\IEEEauthorrefmark{1}\IEEEauthorrefmark{2}, 
        Dan Meng\IEEEauthorrefmark{1}\IEEEauthorrefmark{2}, 
        Mingzhe Zhang\IEEEauthorrefmark{1}\thanks{\textbf{Corresponding Author:} Mingzhe Zhang (zhangmingzhe@iie.ac.cn). This work is supported in part by National Natural Science Foundation of China
        (Grant No.62002339 and No.62125208), the Strategic Priority Research
        Program of the Chinese Academy of Sciences (Grant No. XDB44030200), and
        Joint Funds for Smart Computing of Natural Science Foundation of Shandong
        Province (Grant No.ZR2019LZH014).}\\
        \IEEEauthorrefmark{1} State Key Laboratory of Information Security, Institute of Information Engineering, CAS, Beijing, China. \\
        \IEEEauthorrefmark{2} School of Cyber Security, University of Chinese Academy of Sciences, Beijing, China. \\
        \IEEEauthorrefmark{3} School of Information Science and Engineering, Shandong Normal University, Jinan, China. \\
        damionfan@163.com, \{wangzhiwei, hourui, mengdan, zhangmingzhe\}@iie.ac.cn, xuweizhi@sdnu.edu.cn
}

\begin{document}
\maketitle
\thispagestyle{plain}
\pagestyle{plain}


\begin{abstract}

In the cloud computing era, privacy protection is becoming pervasive in a broad range of applications (e.g., machine learning, data mining, etc). Fully Homomorphic Encryption (FHE) is considered the perfect solution as it enables privacy-preserved computation on untrusted servers. Unfortunately, the prohibitive performance overhead blocks the wide adoption of FHE (about $10,000\times$ slower than the normal computation). As heterogeneous architectures have gained remarkable success in several fields, achieving high performance for FHE with specifically designed accelerators seems to be a natural choice. Until now, most FHE accelerators 
have focused on efficiently implementing one FHE operation at a time based on ASIC and with significantly higher performance than GPU and FPGA. However, recent state-of-the-art FHE accelerators rely on an expensive and large on-chip storage and a high-end manufacturing process (i.e., 7nm), which increase the 
cost of FHE adoption.

In this paper, we propose {\SolutionName}, an FHE acceleration solution based on GPGPU for real applications on encrypted data. {\SolutionName} utilizes Tensor Core Units (TCUs) to boost the computation of Number Theoretic Transform (NTT), which is the part of FHE with highest time-cost. Moreover, {\SolutionName} focuses on performing as many FHE operations as possible in a certain time period rather than reducing the latency of one operation. Based on such an idea, {\SolutionName} introduces operation-level batching to fully utilize the data parallelism in GPGPU. 
We experimentally prove that it is possible to achieve comparable performance with GPGPU as with state-of-the-art ASIC accelerators. {\SolutionName} performs 913 KOPS and 88 KOPS for NTT and HMULT (key FHE kernels) within NVIDIA A100 GPGPU, which is \Tofill{2.61$\times$} faster than state-of-the-art FHE implementation on GPGPU; Moreover, {\SolutionName} provides comparable performance to the ASIC FHE accelerators, which makes it even \Tofill{2.9$\times$} faster than the F1+ with a specific workload. Such a pure software acceleration based on commercial hardware with high performance can open up usage of state-of-the-art FHE algorithms for a broad set of applications in real systems.

\end{abstract}

\section{Introduction}\label{sec:intro}
\begin{figure}[t]
    \centering
    \includegraphics[width=0.75\linewidth]{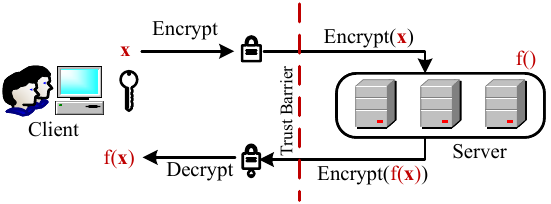}
    \vspace{-10pt}
    \caption{The conceptual workflow of FHE.}
    \label{fig:fhe_concept}
    \vspace{-20pt}
\end{figure}

\vspace{-5pt}
As cloud computing servers are becoming the infrastructure of the world, their security is gaining increasing attention. However, conventional techniques can hardly mitigate all kinds of attacks since defense technology always lags behind new attacking methods. Fortunately, \textit{Fully Homomorphic Encryption (FHE)} provides a new perspective for cloud computing security: it directly processes encrypted data so that attackers cannot get any sensitive data, even with a successful invasion of the system. As shown in Figure \ref{fig:fhe_concept}, the user encrypts and uploads the data to the cloud server, while the server fulfills the computation directly on the encrypted data. Then, the user decrypts the downloaded data to get the results. Compared with other privacy-preserving computing technologies (e.g., Federal Learning\cite{konevcny2015federated}), FHE is general enough to implement different kinds of applications, such as machine learning\cite{lstm,resnet20,helr}, information retrieval\cite{yi2012single} and genome analysis\cite{kim2015private}.

However, the high performance overhead blocks the widespread adoption of FHE in real systems. For example, even with highly optimized FHE libraries and high-end CPUs, FHE computations still cost $10^4\times$ to $10^5\times$ more time than equivalent applications based on undecrypted data.

To bridge this performance gap, a series of optimization solutions have been proposed based on different kinds of hardware. 100x\cite{100x} is the first high-performance CKKS implementation on GPGPU, which supports \texttt{Bootstrap} operations. Unfortunately, it suffers from a lack of hardware support for the modulo calculations. HEAX\cite{heax} is proposed to accelerate CKKS on FPGA, which effectively improves the performance of NTT and modulo operations. However, due to the limited on-chip resources, HEAX can only support a small workload that does not require \texttt{Bootstrap} operations. To meet the requirements of a large scale workload, a series of ASIC accelerators have been proposed, including the F1+\cite{f1}, the CraterLake\cite{craterlake}, the BTS\cite{bts} and the ARK\cite{ark}. These accelerators significantly improve the performance of all FHE operations, which makes it possible to utilize FHE in real workloads. However, all of these works rely on the over-expensive large-scale on-chip buffer or register file (RF), which impedes the popularity of these accelerators. For example, F1+\cite{f1} and CraterLake\cite{craterlake} require 256MB on-chip RFs, while BTS\cite{bts} and ARK\cite{ark} use 512MB RFs.
 
Since high-end GPGPUs have been widely equipped in the cloud servers, it will be very attractive if FHE acceleration can be effectively performed on GPGPUs: on the one hand, using GPGPU to accelerate FHE operations requires very limited economic cost; on the other hand, since most cloud computing tasks (e.g., machine learning, data analysis, etc.) are fulfilled on GPGPUs, a GPGPU-based FHE acceleration solution helps to simplify its integration with other applications. In this paper, we first analyze the 
fundamental reason that limits FHE performance on GPGPU. 
Then, we propose three techniques to improve the performance of all kinds of FHE operations on GPGPU: 1) algorithm optimization for NTT (key kernel of FHE) to improve its hardware efficiency on the GPGPU; 2) NTT optimization for fulfilling the NTT kernels on the emerging TCU hardware; 3) data layout optimization for improving the throughput of the batching FHE operations. 

We evaluate our proposed {\SolutionName} with real workloads and provide a comparison with previous works on CPU, GPGPU, FPGA and ASIC accelerators. The evaluation results show that {\SolutionName} provides significant performance improvement for the key kernel and the operations, which is a speedup of up to \Tofill{397.1$\times$} and \Tofill{1035.8$\times$} for \textit{HMULT} and \textit{HADD}, respectively. When considering overall performance for real workloads, {\SolutionName} achieves higher performance than the F1+\cite{f1} on \Tofill{\texttt{LR} application}, although slightly lower performance than other accelerators. Considering the implementation cost, we believe that the {\SolutionName} is a more competitive solution for accelerating FHE in the cloud server. 

In summary, our contributions are as follows:
\begin{itemize}[wide, labelwidth=!, labelindent=0pt,noitemsep]
    \item We analyze the fundamental reasons from the micro-architectural level for the low performance of NTT, and then propose an optimization at the software level. 
    \item We propose a novel algorithm optimization that allows high-accuracy-performing NTT on the TCU with limited accuracy hardware support.
    \item We optimize the data layout to fully utilize the potential data parallelism for batching FHE operations.
    \item We provide a detailed evaluation of {\SolutionName} from different perspectives. The evaluation results show that our {\SolutionName} provides significantly higher performance than previous works on CPU, GPU and FPGA. Moreover, {\SolutionName} achieves comparable or even higher performance when compared to ASIC accelerators.
\end{itemize}

\section{Background}\label{sec:background}
In this section, we first introduce the basics of \emph{Number Theoretic Transform (NTT)}, which is the key module of FHE. Then, taking CKKS as an example, we briefly introduce the main concepts of FHE. Last, we briefly introduce the micro-architecture of TCU.
For ease of understanding, we summarize all of the symbols used in this paper in Table \ref{tab:ckksnotaion}.

\begin{table}[t]
    \centering
    \caption{Symbols and Notions used in this paper.}\label{tab:ckksnotaion}
    \vspace{-5pt}
    \begin{tabular}{|l|l|l|}\hline
    \textbf{Symbol} & \textbf{Definition} \\ \hline
    Q& (Prime) moduli product $\prod_{i=0}^{L}q_i$\\ \hline
    P& Special (prime) moduli product $\prod_{k=0}^{K}p_k$ \\ \hline
    L& Maximum (multiplicative) level \\ \hline
    dnum& Decomposition number~\cite{han2020better}\\ \hline
    K& Number of special prime moduli \\ \hline 
    N& Degree of a polynomial\\ \hline
    $q_l$& (Prime) moduli, where $0\leq l < L$ \\ \hline
    $p_k$& Special (prime) moduli, where $0\leq k < K$ \\ \hline
    $\psi$, $\psi^{-1}$& Root of unity of twiddle factor for (I)NTT \\ \hline
    \end{tabular}
    \vspace{-15pt}
\end{table}

\subsection{Number Theoretic Transform}\label{sec:background:ntt}

\emph{Number Theoretic Transform (NTT)} is a specialized form of \emph{Discrete Fourier Transform (DFT)} and widely adopted in various cryptography applications. Unlike DFT, NTT uses the $\psi$ as the primitive $N$-th root of unity to convert polynomials in a finite field of integers, where $\psi^{N}_{N} \equiv 1$ {\rm mod} $q$ for a given $N$ and a prime $q$. According to Fermat’s Little Theorem\cite{fermatlittle}, for the prime $q$, exists at least one primitive root $g$ such that $g^{(q-1)} \equiv 1$ {\rm mod} $q$, generating the primitive $N$-th root of unity $\psi_{N}=g^{(q-1)/N}$ {\rm mod} $q$. Therefore, the overall NTT algorithm can be formulated as
\vspace{-5pt}
\begin{equation}
    \vspace{-5pt}
    A_{k}=\sum_{n=0}^{N-1}(a_{n}\psi_{(N,q)}^{nk} \mod q), k\in[0,N)
\end{equation}
where $a_n$ indicates the $n$-th coefficient of the input polynomial, $A_{k}$ indicates the $k$-th coefficient of the output polynomial, and $\psi_{(N,q)}$ indicates the primitive $N$-th root of the NTT's unity for $Z_{q}$. Similarly, the \emph{inverse NTT (INTT)}, which is the inverse process of NTT, can be formulated as
\vspace{-5pt}
\begin{equation}
    \vspace{-5pt}
    a_{k}=\frac{1}{N} \sum_{n=0}^{N-1}(A_{n}\psi_{(N,q)}^{-nk} \mod q), k\in[0,N)
\end{equation}
where $\psi_{(N,q)}^{-1}$ refers to the primitive $N$-th root of the INTT's unity for $Z_q$, and $\psi_{N}^{-1}=\psi_{N}^{(q-2)}$ in the light of Fermat's Little Theorem\cite{fermatlittle}.

Based on NTT, the polynomial multiplication of $A(X)$ and $B(X)$ can be performed as $A(X)\cdot B(X)=INTT(NTT(A(X))\odot NTT(B(X)))$. To avoid applying the NTT on a $2N$-length input with $N$ zero-padding, a negative-cyclic convolution is used to maintain the multiplication in a polynomial ring $Z[X]/(X^{N}+1)$ \cite{poppelmann2012towards}. The coefficients of the 
result in the polynomial ring are the same as the output of the negative-cyclic convolution. Therefore, the polynomial multiplication between $A(X)$ and $B(X)$ can be formulated as
\begin{equation}
    \begin{split}
        &c = \psi^{-1}\odot INTT(NTT(\bar{a})\odot NTT(\bar{b})), \\
        &\Psi^{-1}=\{1,\psi_{(2N,q)}^{-1},\psi_{(2N,q)}^{-2},...\psi_{(2N,q)}^{-(N-1)}\}
    \end{split}
\end{equation}
where the operator $\odot$ indicates the element-wise multiplication of the coefficient of two polynomials, the $\bar{a}$, $\bar{b}$ and $\bar{c}$ are the vectors composed by the coefficients of $A(\psi_{2N,q}\cdot X)$, $B(\psi_{2N,q}\cdot X)$ and $C(X)$. By merging the $\psi_{(2N,q)}$ and the $\Psi^{- 1}$ with NTT/INTT, the new formula integrated with negative-cyclic convolution can be represented as
\vspace{-5pt}
\begin{equation}
    \vspace{-5pt}
    A_{k} = \sum_{n=0}^{N-1}((a_{n}\psi_{2N}^{2nk+n})\bmod q).\label{eq:ntt-negative-cyclic}
\end{equation}

\begin{figure}
    \centering
    \includegraphics[width=0.75\linewidth]{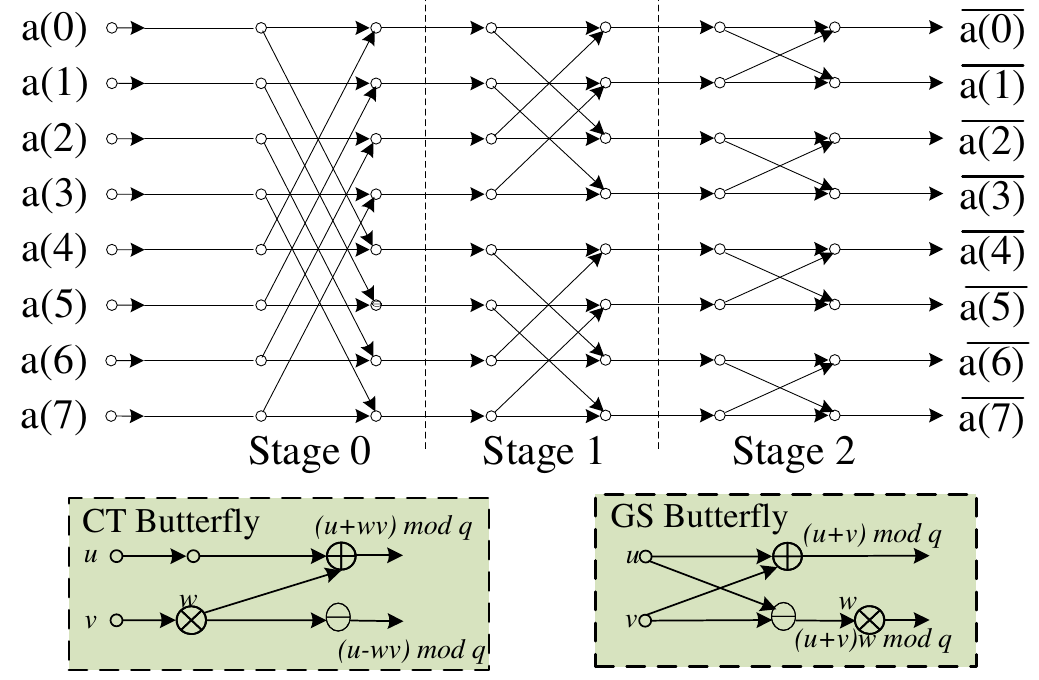}
    \vspace{-5pt}
    \caption{An example of 8-point polynomial NTT/INTT with butterfly operation. Especially, the \textit{CT Butterfly} is used in NTT, and the \textit{GS Butterfly} is for INTT.}
    \label{fig:butterfly}
    \vspace{-15pt}
\end{figure}

As shown in Eq. \ref{eq:ntt-negative-cyclic}, the integer modular multiplication operation, which is the basic computation operation of NTT/INTT, produces high computational overhead\cite{kim2020accelerating}. To tackle this issue, the Cooley-Tukey \cite{cooley1965algorithm} and Stockham\cite{cochran1967fast} algorithms are leveraged to reduce the computation complexity from $O(N^{2})$ to $O(Nlog_{N})$, which reduces the amount of integer modular multiplication operation in NTT. 

As shown in Figure \ref{fig:butterfly}, the Cooley-Tukey and Stockham algorithms use \emph{butterfly operation} to calculate the two elements of interval $N/(2i)$ at $stage_{i}$, and the later $stage_{(i+1)}$ depends on the result produced by the prior $stage_{i}$, which causes the data-dependency issue between the neighboring stages. 
Prior works \cite{chen2022cfntt,dai2015cuhe,durrani2021accelerating,kim2020accelerating} exploit the \emph{split-radix} technique to relieve the dependency issues, which divide the $N$-point of NTT into $k$ interleaved $N/k$-point NTTs (also known as the \emph{radix-$k$ NTT}). However, the radix-$k$ NTT can hardly remove all dependency issues and inevitably requires extra synchronizations. In this work, we 
eliminate dependency issues by fulfilling the NTT algorithm with matrix multiplications.

\subsection{CKKS Scheme}\label{sec:background:ckks}

CKKS is an emerging FHE scheme with support for fixed-point real number\cite{ckks}, which is considered to be one of the most prominent FHE schemes for real-world tasks\cite{lstm,helr,resnet20,resentop}. In this paper, we focus on CKKS, but the proposed technique can also be applied to other FHE schemes, such as BFV\cite{bfv} and BGV\cite{BGV}. We will discuss the generality of our work in Section \ref{sec:discussion}.

In CKKS, each $N/2$ input real numbers are encoded as
\vspace{-5pt}
\begin{equation}
    \vspace{-5pt}
    m(X)=\sum_{i=0}^{N-1}c_{i}X^{i},~N=2^n,~n\in[10,18]
\end{equation}
where $m(X)$ is a polynomial in plaintext with the cyclotomic polynomial ring $\mathbb{R}_{Q}=\mathbb{Z}_{Q}[X]/(X^{N}-1)$. Each $m(X)$ contains $N$ coefficients (\{$c_i$\}), which are integers moulded by Q. Q is a prime number with hundreds or even thousands of bits that relate directly to the HE ciphertext spaces. Then CKKS encrypts the $m(x)\in \mathbb{R}_Q$ into a ciphertext polynomial pair ($a(X),b(X)$) as
\begin{equation}
    \vspace{-5pt}
    \begin{split}
        ct &= (a(X),b(X))\\ 
        &= (b(X)\cdot s(X)+m(X)+e(X),b(X))
    \end{split}
\end{equation}
where $ct$ refers to the ciphertext, $s(X)\in R_Q$ refers to the secret key, $a(X)\in R_Q$ refers to the random polynomial and $e(X)$ refers to the small \emph{Guassian Error polynomial} that guarantees the LWE security\cite{ckks}. Based on the ciphertext, the FHE operations in CKKS can be summarized as follows:
\begin{itemize}[wide, labelwidth=!, labelindent=0pt,noitemsep]
    \item \texttt{HMULT(A,B)} fulfills the multiplication of two ciphertexts ($A(a_0(X), b_0(X))$ and $B(a_1(X), b_1(X))$) as ($a_0\cdot a_1$,~$a_0\cdot b_1 + a_1\cdot b_0$)+\texttt{KeySwitch}($b_0\cdot b_1$).
    \item \texttt{CMULT(A,B)} fulfills the multiplication between ciphertext A($a_0(X),b_0(X)$) and plaintext B($a_1(X)$) in the manner of element-wise multiplication, which can be formalized as ($a_0(X)\cdot a_1(X)$, $b_0(X)\cdot a_1(X)$).
    \item \texttt{HADD(A,B)} adds the ciphertext A($a_0(X), b_0(X)$) to the other ciphertext B($a_1(X), b_1(X)$) in the manner of element-wise addition, which can be formalized as A($a_0(X),~b_0(X)$)$+$B($a_1(X),b_1(X)$)=($a_0(X)+a_1(X)$, $b_0(X)+ b_1(X)$).
    \item \texttt{HROTATE(A,\textit{r})} circularly shifts the $r$-th ciphertext with the granularity of element-wise, which usually serves the accumulative operation of ciphertexts. \texttt{HROTATE(A, \textit{r})} can be formalized as A($0,~rotate(b_0,r)$)+ \texttt{KeySwitch}($rotate(a_0,r)$).
    \item \texttt{RESCALE(A)} updates the security level budget after the execution of \texttt{HMULT} and \texttt{CMULT}, which can be formalized as A($rescale(a_0),rescale(b_0)$).
\end{itemize}

The most significant performance bottleneck of CKKS is the overly high computational overhead. To tackle this issue, the following techniques are applied:
\begin{itemize}[wide, labelwidth=!, labelindent=0pt,noitemsep]
    \item \textbf{Residue Number System (RNS).} To enable modulo computation of the wide Q and coefficients, Chinese Remainder Theorem (CRT) is usually used to decompose the coefficients, which introduces huge computational overhead~\cite{erabelli2020pyfhe}. Therefore, the Residue Number System (RNS)~\cite{fullrnsckks} is represented to convert the ciphertext polynomial with the wide coefficients to \textit{L} residue polynomials with 32-bit coefficients, each polynomial coefficient computed as $c_{i}\mod q_{l}$, where $0\leq i<N$ and $0\leq l<L$. In this way, the Full-RNS CKKS scheme significantly reduces the 
    complexity of FHE operations~\cite{bajard2016a}.
    \item \textbf{Generalized Key-Switching (GKS).} The state-of-the-art GKS technique~\cite{han2020better} reduces computational cost by balancing $L$, which decomposes the Q into $dnum$ slices ($\{Q_{j}\}_{0\leq j<dnum}$=\{$\prod_{i=j\alpha}^{(j+1)\alpha -1}q_{i}\}_{0\leq j<dnum}$, where $\alpha=(L+1)/dnum$). GKS allows that P=$\prod_{k=0}^{K-1}p_{k}$ only needs to be larger than $MAX_{0\leq j<dnum}(Q_j)$. Thus, the ciphertext computed in \texttt{keySwitch} 
   can be decreased by adjusting $dnum$. 
\end{itemize}

In this paper, we adopt Full-RNS together with generalized key-switching in our scheme to pursue the best performance. 

\vspace{-5pt}
\subsection{Tensor Core Unit (TCU)}\label{sec:background:tcu}

\begin{figure}[t]
    \centering
    \includegraphics[width=0.9\linewidth]{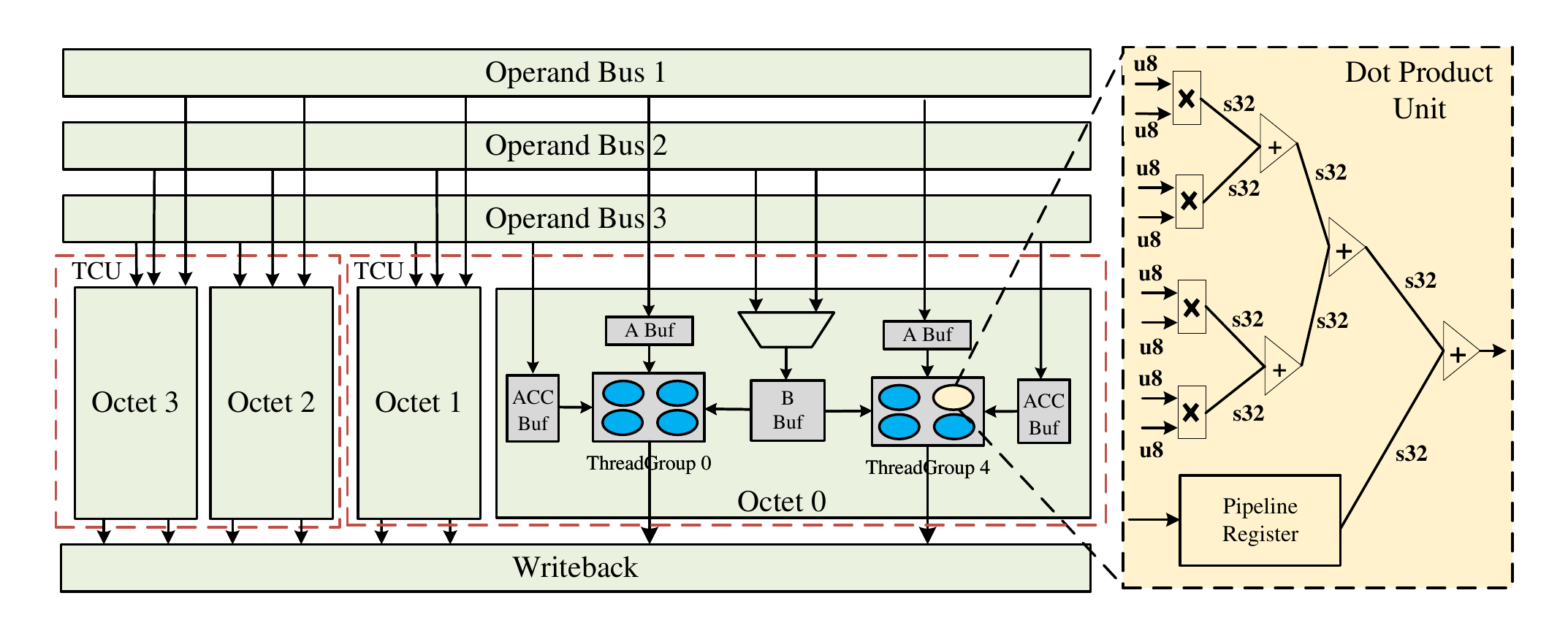}
    \vspace{-15pt}
    \caption{Illustration of the TCU micro-architecture.}\label{fig:tcu}
    \vspace{-15pt}
\end{figure}

The TCU is a specialized component for accelerating \emph{multiply and accumulate (MAC)} operations. It was first introduced to NVIDIA GPGPUs in the Volta architecture\cite{v100} and then improved in the Turing architecture\cite{turing} and the Ampere architecture\cite{a100}. As shown in Figure \ref{fig:tcu}, each TCU is composed of two octets, while one warp simultaneously uses two TCUs. In one octet, each thread group owns a specific buffer for the LHS matrix (A Buf) and accumulation results (ACC Buf). The two thread groups in the same octet share one RHS matrix buffer (B Buf) and select its source with a multiplexer.

Although a TCU achieves significantly improved performance\cite{v100}, it can only support the low-precision computation (i.e., \texttt{FP16}, \texttt{BF16}, \texttt{INT8}, \texttt{INT4} and \texttt{INT1}). Taking the \texttt{UINT8}  MAC as an example, each thread can access a four-by-four \emph{Dot Product Unit (DPU)}, and each multiplexer in the DPU generates and sends the production to a \texttt{SINT32} accumulator for further computation. Due to the limited precision, TCUs cannot be directly utilized for FHE acceleration. This is the first work to accelerate FHE operations on TCUs.  


\section{Motivation}\label{sec:motiv}
In this section, we provide a comprehensive understanding of running FHE on GPGPU at the micro-architecture level. We first quantitatively analyze the underutilization of hardware. 
Then, we present the challenges of using the emerging TCUs in FHE acceleration. Finally, we summarize the opportunities for a pure software solution to accelerate the FHE on GPGPUs.  

\vspace{-5pt}
\subsection{Inefficient NTT Computation}\label{sec:motiv:ntt}

\begin{figure}[t]
    \centering
    \includegraphics[width=0.9\linewidth]{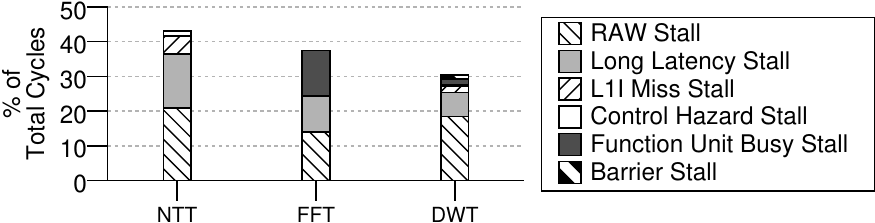}
    \vspace{-5pt}
    \caption{GPGPU pipeline-stall breakdown of different algorithms based on \emph{Butterfly} operations. The thread block sizes of NTT, FFT and DWT are 128, 192 and 256, respectively.}\label{fig:motiv1}
    \vspace{-10pt}
\end{figure}


To investigate the pipeline stall issue of NTT, we simulate an NVIDIA 1080Ti GPU on GPGPUSim\cite{khairy2020accel} and run a state-of-the-art NTT implementation\cite{dai2015cuhe} on it. We also run FFT and DWT implementations from the Rodinia Benchmark Suite\cite{che2009rodinia}.
Note that, since the GPGPU switches to the other warp when a pipeline stall occurs, here we consider only the stall cycles that cannot be hidden. The results are shown in Figure \ref{fig:motiv1}, and we can make the following observations:
\begin{itemize}[wide, labelwidth=!, labelindent=0pt,noitemsep]
    \item All kernels suffer from the pipeline stall. Especially, the proportion of the pipeline stall time is up to \Tofill{43.2\%} for NTT. 
    \item Not surprisingly, the most significant reason for the pipeline stall is the \emph{Read-After-Write (RAW)} issue in all kernels, which can be inferred as the result of data-dependency between the neighboring stages in the \emph{butterfly operations}. For the NTT kernel, the proportion of the RAW stall is \Tofill{20.9\%}, which is \Tofill{48.6\%} of its overall pipeline stalls.
\end{itemize}

Therefore, the key to boosting NTT on GPGPU is to reduce the RAW stall as much as possible.
\vspace{-5pt}
\subsection{Low Computation Resource Utilization}\label{sec:motiv:low-utilization}
\begin{figure}[t]
    \centering
    \includegraphics[width=0.9\linewidth]{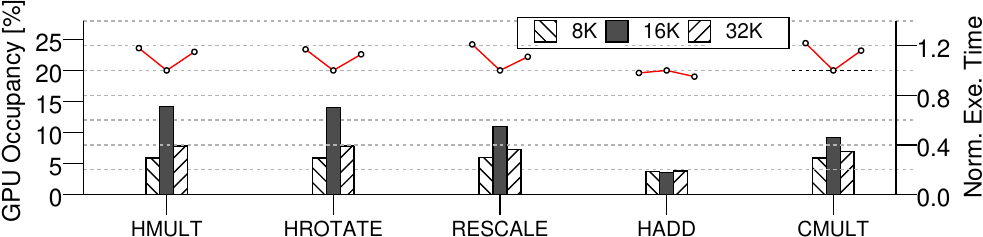}
    \vspace{-5pt}
    \caption{\RebuttalChange{Impact of threading on the GPGPU occupancy and performance for CKKS operations. The bars indicate the GPU occupancy (left), and the line charts indicate the Normalized Execution Time (right).}}\label{fig:motiv2}
    \vspace{-15pt}
\end{figure}

As GPGPU utilizes a large amount of SIMD cores with the static pipeline, once a pipeline stall occurs, the scheduler directly switches to other pending threads to fully utilize the hardware resources. Therefore, increasing the thread number for each Streaming Multiprocessor (SM) would help to improve the hardware occupancy by better hiding the possible stalls. However, as the thread in one SM increases, the performance will be hurt due to there being more intense competition for the resources inside the SM.

To investigate this tradeoff, we run the instance of TensorFHE-NT as described in Table \ref{tab:designs} with no batching on the NVIDIA A100 GPGPU (see Table~\ref{tab:platform_config} for detail) and monitor the hardware occupancy and the operation performance. \RebuttalChange{We increase the total number of threads from 8196 (8K) to 32768 (32K) to show its impact. Note that, we evaluate different thread number per SM and report the best performed results, which contains 512 threads in each SM.} 
As shown in Figure \ref{fig:motiv2}, we can make the following observations:
\begin{itemize}[wide, labelwidth=!, labelindent=0pt,noitemsep]
    \item \RebuttalChange{Not surprisingly, as the thread number increases from 8K to 16K, the occupancy of all CKKS operations grows. In particular, for the most frequently used \texttt{HMULT}, the GPGPU occupancy is up to \Tofill{14.3\%}. Accordingly, the execution time of all operations is also reduced.}    
    \item \RebuttalChange{However, as the thread number continues increasing to 32K, the GPGPU occupancy of all operations degrades. This is because each thread gets fewer data and the total number of memory accesses increases, which leads to less efficient bandwidth utilization. Also, the execution time is increased.} The only exception occurs in \texttt{HADD} since the workload for each thread is minor and the execution time ($<0.1\mu s$) is significantly affected by other factors.
    \item Moreover, the highest GPGPU occupancy for all kinds of operations is lower than $15\%$. When considering the performance, the under utilization of GPGPU is unacceptable: with the best performance of all operations, GPGPU occupancy is around $10\%$ and even lower than $5\%$ for the \texttt{HADD}. 
\end{itemize}

Ideally, we want a system that fully utilizes the hardware resources to achieve performance that is as high as possible. But state-of-the-art CKKS fails on this goal.

\vspace{-5pt}
\subsection{Ineffective Usage of Emerging Hardware}\label{sec:motiv:emerging}

As we introduced in Section \ref{sec:background:tcu}, modern GPGPU uses TCUs to accelerate the MACs and achieves great success with AI workloads\cite{feng2021apnn}. However, the limited precision support makes it hard to accelerate the arithmetic kernels of CKKS (i.e., NTT/INTT) with TCUs. For example, in the state-of-the-art implementation of CKKS, the NTT/INTT kernel is based on the computation for \texttt{INT32}, while the TCUs only support up to \texttt{INT8}. Besides, GPGPU lacks efficient hardware support for the modulo operation, which also significantly affects the performance of CKKS on GPGPU\cite{100x}.

\vspace{-5pt}
\subsection{Our Opportunities}
Based on the above observations, we conclude our opportunities for boosting FHE on GPGPU as follows:
\begin{itemize}[wide, labelwidth=!, labelindent=0pt,noitemsep]
    \item As discussed in Section \ref{sec:motiv:ntt}, the NTT computation suffers from serious RAW stalls and low efficiency of modulo operations. Therefore, we could optimize the NTT algorithm 
    to remove the RAW stalls and the excessive modulo operations.
    \item As discussed in Section \ref{sec:motiv:emerging}, modern GPGPU contains emerging TCUs for high-performance MAC computations. To fully utilize the powerful 
    TCUs, algorithm optimization for NTT is necessary, which aims to apply high-accuracy NTT based on the hardware with limited accuracy support. 
    \item As discussed in Section \ref{sec:motiv:low-utilization}, current thread-level parallelism cannot sufficiently utilize the abundant hardware resources of GPGPU. Therefore, we exploit the overall performance improvement of FHE with batching techniques. \  RebuttalChange{However, Ref \cite{100x} observes that for all CKKS operations, the bandwidth is more intensive than the computation resources, which limits the capability of batching execution. To tackle this issue, we break the CKKS operations into a series of reusable kernels, which balances the bandwidth and computation requirement of each kernel. Such a hierarchical reconstruction model also helps to better understand and optimize the complex CKKS scheme.}
\end{itemize}

Consequently, we propose a series of optimizations and finally integrate them together. This, in turn, improves the performance of FHE execution on GPGPU. In the next section, we introduce our detailed scheme: \emph{{\SolutionName}}.

\section{TensorFHE}\label{sec:design}
Based on the analysis made in Section \ref{sec:motiv}, we propose \emph{TensorFHE}, which accelerates FHE operations on GPGPU. TensorFHE is based on a hierarchical reconstruction of CKKS, which decomposes the CKKS operations into a series of reusable arithmetic kernels. For the most time-consuming NTT/INTT kernels, TensorFHE first uses an optimized algorithm to reduce pipeline stalls. It then further improves the algorithm to exploit accelerating NTT/INTT kernels by using the emerging TCU. For the rest of the kernels, TensorFHE conducts the paralleled algorithms on CUDA cores. Moreover, to fully utilize the potential computational and data-level parallelism of GPGPU, TensorFHE also introduces an optimization scheme for batching multiple FHE operations. As a result, TensorFHE provides significant performance improvement for CKKS on GPGPU with pure software optimizations. In the rest of this section, we introduce TensorFHE in detail.

\subsection{Hierarchical Reconstruction of CKKS}

\begin{table}[t]
    \centering
    \caption{Hierarchical reconstruction model of CKKS.}\label{tab:hierarchy}\vspace{-5pt}
    \begin{tabular}{|l|l|l|}\hline
        \textbf{Operation} & \textbf{Description} & \textbf{Composing Kernels} \\ \hline
        
        \multirow{2}{*}{\centering{\texttt{HMULT}}}
          & \multirow{2}{*}{Multiply two ciphertexts.} & \multirow{1}{*}{\centering{{NTT}, {Hada-Mult},}}  \\ 
          & &{Conv}, {Ele-Add}  \\ \hline
  
          \multirow{2}{*}{\centering{\texttt{CMULT}}}
          & \multirow{1}{*}{Multiply ciphertext with} & \multirow{2}{*}{\centering{{Hada-Mult}, {Ele-Add}}}  \\ 
          &  plaintext.&  \\ \hline
      
        \multirow{2}{*}{\centering{\texttt{HROTATE}}}
          & \multirow{2}{*}{Roatate ciphertext.} & \multirow{1}{*}{\centering{NTT, {Hada-Mult}, {Ele-Add},}}  \\ 
          & &  Conv, ForbeniusMap \\ \hline   

        \multirow{2}{*}{\centering{\texttt{RECALE}}}
          & \multirow{1}{*}{Reduce the security level} & \multirow{2}{*}{\centering{NTT, {Ele-Sub}}}  \\ 
          &of ciphertext. &  \\ \hline       
     
     \multirow{2}{*}{\centering{\texttt{HADD}}}
          & \multirow{2}{*}{Add two ciphertexts.} & \multirow{2}{*}{\centering{{Ele-Add}}}  \\ 
          & &  \\ \hline      
        
    \end{tabular}
    \vspace{-15pt}
\end{table}

To better understand and optimize the CKKS scheme, TensorFHE uses a hierarchical model to reconstruct the CKKS scheme. As shown in Table \ref{tab:hierarchy}, each FHE operation can be decomposed into multiple reusable arithmetic kernels. Overall, there are seven involved kernels as follows:
\begin{itemize}[wide, labelwidth=!, labelindent=0pt,noitemsep]
    \item \emph{\textbf{NTT}} transforms coefficient represented polynomial into point-value representation to accelerate the polynomial multiplication. It can be formulated as $A_k = \sum_{n=0}^{N-1}(x_{n}\psi_{(N,p)}^{nk}\mod q)$, where $0\leq k <N$ \cite{cheon2018full}. 
    \item \emph{\textbf{Hadamard Multiplication (Hada-Mult)}} can be formulated as $c=(a\circ b)\mod q$~\cite{cheon2018full}. The $a\circ b$ indicates the element-wise product of two polynomials represented as vectors (also known as \textit{Hadamard product})\cite{cheon2017homomorphic}.
    \item \emph{\textbf{Element-wise Addition(Ele-Add)}} and \emph{\textbf{Element-wise Subtract (Ele-Sub)}} can be formulated as $c=(a\oplus b)\mod q$ and $c=(a\ominus b)\mod q$, respectively\cite{cheon2018full}. Here we use $a\oplus b$ and $a\ominus b$ to indicate the element-wise addition and subtraction for the two polynomials represented as vectors\cite{cheon2018full}.
    \item \emph{\textbf{ForbeniusMap}} performs the Frobenius map function for the polynomial by index $r$ under the NTT domain \cite{han2020better}. For every $a^{(i)}$=$(a_{j}^{(i)})_{j\in[0,N-1]}$ in $\{a^{(i)}\}_{i\in[0,l]}$, it generates $a^{'(i)}$=$((a^{(i)}_{\pi_{r}^{-1}(j)})_{j\in[0,N-1]}$, where $\pi_{r}(x)$=$([5^{r}(2x+1)]_{2N}-1)/2$, which is a permutation operation. Then, this kernel will return $a^{'}$=$(a^{'(i)})_{i\in[0,l]}$.
    \item \emph{\textbf{Conjugate}} indicates the conjugation of the coefficients in the polynomial with the modulus $q$ \cite{mukherjee2016cyclotomic}. For the polynomial \textbf{A} represented by coefficients, the conjugate representation of polynomial is $\bar{A}$.
    \item \emph{\textbf{Fast basis Conversion}} Kernel (Conv) converts a set of residue polynomials to another set whose prime moduli is different from the former. It is the key operation of \texttt{ModUp}, \texttt{ModDwon} and \texttt{ModRaising} \cite{cheon2018full}. 
\end{itemize}

\begin{figure}[t]
    \centering
    \includegraphics[width=\linewidth]{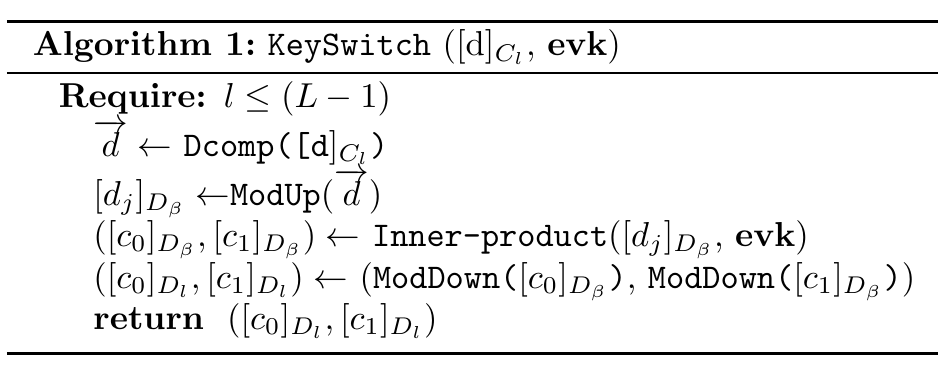}
    \label{alg:keyswitch}
\vspace{-30pt}
\end{figure}

\begin{figure}[t]
    \vspace{-15pt}
    \centering
    \includegraphics[width=\linewidth]{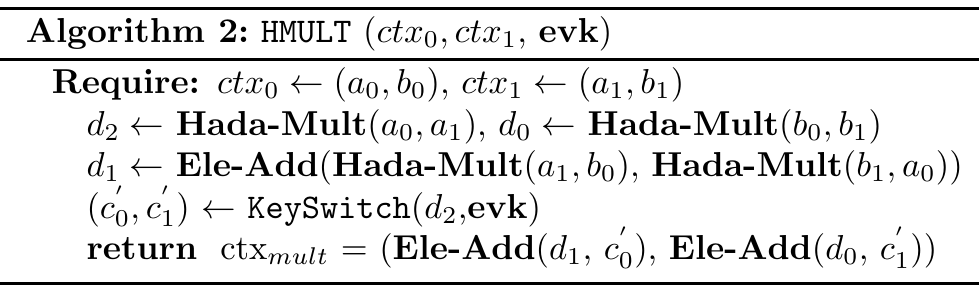}
    \label{alg:hmult}
\vspace{-20pt}
\end{figure}

\begin{figure}[h]
    \vspace{-10pt}
    \centering
    \includegraphics[width=\linewidth]{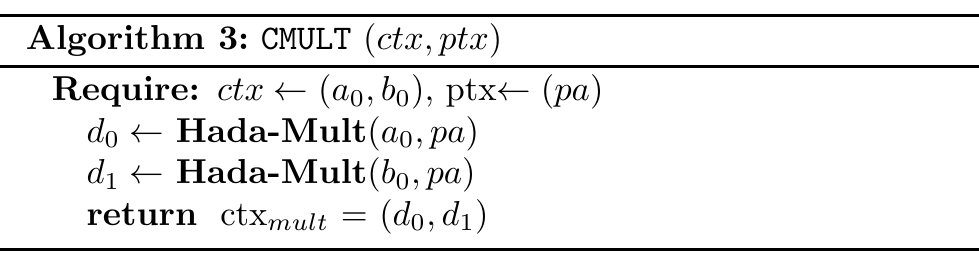}
    \label{alg:cmult}
\vspace{-30pt}
\end{figure}

\begin{figure}[t]
    \vspace{-15pt}
    \centering
    \includegraphics[width=\linewidth]{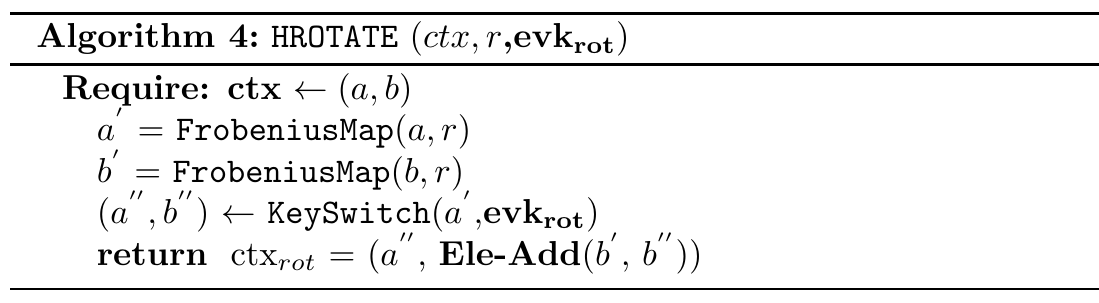}
    \label{alg:hrotate}
\vspace{-20pt}
\end{figure}

\begin{figure}[t]
    \vspace{-10pt}
    \centering
    \includegraphics[width=\linewidth]{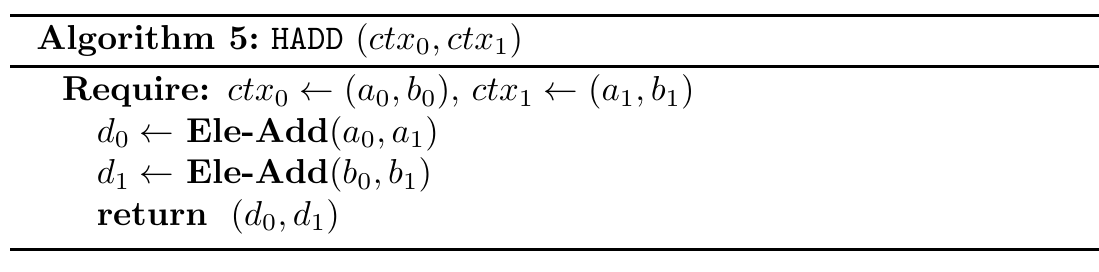}
    \label{alg:hadd}
\vspace{-25pt}
\end{figure}

\begin{figure}[t]
    \vspace{-15pt}
    \centering
    \includegraphics[width=\linewidth]{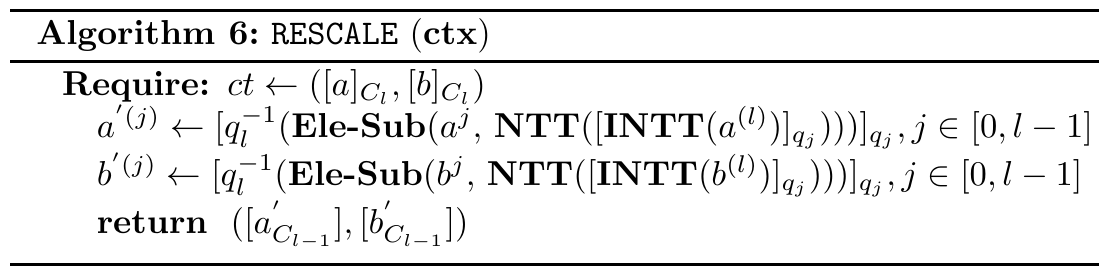}
    \label{alg:rescale}
\vspace{-5pt}
\end{figure}

\begin{figure}[h]
    \vspace{-15pt}
    \centering
    \includegraphics[width=\linewidth]{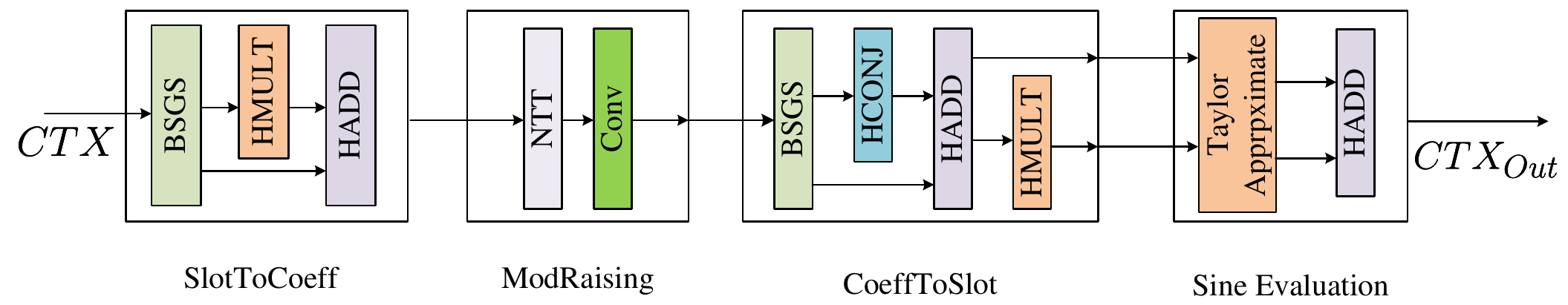}
    \vspace{-20pt}
    \caption{\texttt{Bootstrap} workflow composed by various kernels.}\label{fig:bootstage}
\vspace{-10pt}
\end{figure}

Based on the above arithmetic kernels, all CKKS operations can be composed as follows:
\begin{itemize}[wide, labelwidth=!, labelindent=0pt,noitemsep]
    \item \textbf{\texttt{keySwitch}} refreshes the secret key of the ciphertext and maintains its precision, which is widely adopted in several FHE operations. As shown in Alg. 1, the \texttt{keySwitch} mainly includes the \texttt{ModUp}, the \texttt{ModDown}, the \texttt{Dcomp} and the \texttt{Inner-product}. For the \texttt{ModUp} and the \texttt{ModDown}, we directly use the method introduced in Ref \cite{han2020better}, which can be composed by multiple \emph{NTT}s and \emph{Conv}s. For the \texttt{Dcomp}, which represents the RNS decomposition operation, it can be implemented with minor modified \emph{Conv}\cite{100x}. \texttt{Inner-product} is the key operation in \texttt{KeySwitch}; it is implemented with multiple \emph{Hada-Mult}s and \emph{Ele-Add}s.
    \item \textbf{\texttt{HMULT}} and \textbf{\texttt{CMULT}} can be implemented with \emph{Hada-Mult} and \emph{Ele-Add} (as shown in Alg. 2 and 3). Note that the \texttt{HMULT} requires \texttt{KeySwitch} to reduce the ciphertext precision loss.
    \item \textbf{\texttt{HROTATE}} is composed by \emph{ForbeniusMap} and \emph{Ele-Add}, which also needs \texttt{Keyswitch} (as shown in Alg. 4).
    \item \textbf{\texttt{HADD}} requires only \textbf{Ele-Add} kernels (as shown in Alg. 5). 
    \item \textbf{\texttt{RESCALE}} consists of multiple \emph{NTT} and \emph{Ele-Sub} (Alg. 6). 
    \item \textbf{\texttt{Bootstrap}} includes four stages: the \texttt{SlotToCoeff}, the \texttt{ModRaising}, the \texttt{CoeffToSlot} and the \texttt{Sine Evaluation} (as shown in Figure \ref{fig:bootstage}). In this work, the \texttt{Bootstrap} is implemented according to the \textit{slim botstrapping}\cite{chen2018homomorphic}, which reorders the stages and reduces the computational overhead of the \texttt{SlotTocoeff}. In the \texttt{SlotToCoeff} and \texttt{CoeffToSlot} stages, we use the \emph{Baby-Step Giant-Step (BSGS)}\cite{shoup1995new} algorithm to fulfill the DFT computation, which is the most time-consuming operation and composed by multiple \texttt{CMULT}, \texttt{HMULT} and \texttt{HROTATE} operations. To improve BSGS performance, we adopt the \emph{Faster Homomotrphic DFT algorithm}\cite{cheon2018faster}, which significantly reduces the requirement for the \texttt{HROTATE} and the \texttt{CMULT}. Besides, we use the \emph{Taylor Polynomial Approximation}\cite{brunelli2009approximating} to approximate the \emph{sin(x)} function\cite{han2020better}.
\end{itemize}

Based on this reconstruction, {\SolutionName} accelerates the CKKS operations with large-scale parallelism on GPGPU for all kernels. Especially for the NTT/INTT, which is the most time-consuming kernel\cite{al2018faster,mert2019design}, {\SolutionName} exploits to accelerate it with the CUDA cores and the TCUs, which will be introduced in the rest of this section.

\subsection{NTT Optimization for CUDA Core}

According to Section \ref{sec:background:ntt}, the NTT algorithm can be formulated as Eq. \ref{eq:ntt-negative-cyclic} and implemented in the form of the \emph{butterfly algorithm}\cite{longa2016speeding}. However, the performance of the \emph{butterfly algorithm} suffers from the frequent pipeline stalls caused by the RAW dependency issue. Besides, the large amount of modulo operations in Eq.\ref{eq:ntt-negative-cyclic} also consumes a large amount of time due to the GPGPU's lack of efficient hardware support for modular arithmetic.

To overcome these issues, we use a transform technique to optimize NTT. Considering the modular arithmetic's compatibility with addition, Eq. \ref{eq:ntt-negative-cyclic} can be transformed as 
\vspace{-5pt}
\begin{equation}
\vspace{-5pt}
    A_k = \left ( \sum_{n=0}^{N-1}a_{n}\psi_{(2N,q)}^{2nk+n} \right ) \mod q, a_n \in a. \label{eq:ntt_fusion}
\end{equation}

Such transformation converts the NTT from the butterfly algorithm to matrix-vector multiplication as
\begin{equation}
    A=(W_{N\times N} \times a^{T}) \mod q, w_{ij}=\psi_{(2N,q)}^{2ij+j}\in W_{N\times N}.  \label{eq:ntt_mvm}
\end{equation}


Such transformation can benefit the NTT performance in the following aspects: 
\begin{itemize}[wide, labelwidth=!, labelindent=0pt,noitemsep]
    \item \textbf{Data Reuse.} According to the CKKS definition, the \emph{twiddle factor matrix} is determined by the FHE parameters, such as $N$ and $q$. Therefore, for one CKKS instance, the \emph{twiddle factor matrix} $W_{N\times N}$ can be pre-computed in the initialization and reused by all NTT operations. This can significantly reduce the computational and storage overhead.
    \item \textbf{Hardware Efficiency.} The matrix-vector multiplication has good memory locality and can be easily paralleled on GPGPU with high pipeline efficiency.
    \item \textbf{Modulo Reduction.} Only one modulo operation is required for each $A_k$, and a large amount of time can be saved due to the modulo reduction. Note that, since we are reducing the number of modulo operations, more memory space is required to store the longer temporary values. In this paper, we use a 64-bit integer as the accumulative variable, which does not meet the overflow issue until $N \leq 2^{18}$.
\end{itemize}

However, since the length of the polynomial ($N$) usually ranges from $2^{10}$ to $2^{18}$, the scale of the \emph{twiddle factor matrix} ($W_{N\times N}$) and the input vector ($a$) will be extremely large, which leads to over-high computation and memory overhead. To tackle this issue, we adopt the Cooley-Tukey Recursive algorithm~\cite{gupta1990fast}, which transforms the input vector ($a$) to an input matrix ($a_{N_1\times N_2}$, where $N=N_1\times N_2$). Accordingly, the scale of the \emph{twiddle factor matrix} is also reduced to $N_1\times N_1$. In this way, the NTT can be further transformed as 
\vspace{-5pt}
\begin{equation}
    \mathbf{A}=((\mathbf{a_{N_1\times N_2}}\times\mathbf{W_1})^{T}\odot \mathbf{W_2})\times \mathbf{W_3}\mod q \label{eq:ntt_gemm}
\end{equation}
where the scale of $W_1$, $W_2$ and $W_3$ is $N_1\times N_1$, $N_1\times N_2$ and $N_2\times N_2$, respectively. Especially, the elements of the three matrices can be represented as $\psi_{(2N_1,q)}^{2ij+j}$, $\psi_{(2N,q)}^{2ij+j}$ and $\psi_{(2N_2,q)}^{2ij}$. 

In this way, the complex NTT is transformed to three sequential matrix-matrix multiplications. Although the time complexity increases, the transformed algorithm could still benefit from better utilization of parallelism.

\subsection{NTT Optimization for Tensor Core}

We further exploit to accelerate the NTT kernel with the emerging TCUs, which is considered to be the most powerful unit for MACs in the GPGPU. As mentioned in the above subsection, the NTT kernel is transformed into three sequential matrix multiplications, which provides the possibility of using TCUs. However, all matrix elements used in the above mentioned NTT kernel are UINT32, while TCU supports only the low-precision arithmetic operations (i.e., up to 8-bits for integer and 16-bits for floating-point data). We use a \emph{segment-fusion} scheme to fulfill the GEMM without precision loss to overcome this issue.

\begin{figure}[t]
    \centering
    \includegraphics[width=\linewidth]{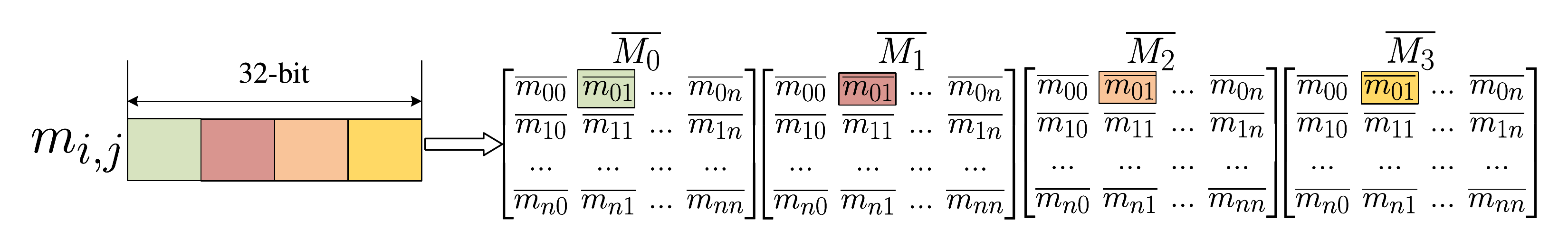}
    \vspace{-20pt}
    \caption{Illustration of the segmentation process for the matrix based on a 32-bit integer.}\label{fig:U32ToU8}
    \vspace{-20pt}
\end{figure}

Figure \ref{fig:U32ToU8} presents the segmentation process. For each 32-bit integer element $m_{i,j}$ in the matrix $M$, we segment every consecutive 8-bits into an 8-bit integer. Then, each 8-bit integer is distributed to the corresponding place of $m_{i,j}$ in the segmented matrices. The matrix selection is according to the original location of the 8-bit integer. In this way, we can get four matrices composed by 8-bit elements ($\overline{M_0}$, $\overline{M_1}$, $\overline{M_2}$, $\overline{M_3}$), which can be processed by the TCUs.  


\begin{figure*}[t]
    \centering
    \begin{minipage}[l]{0.65\linewidth}
\includegraphics[width=1\linewidth]{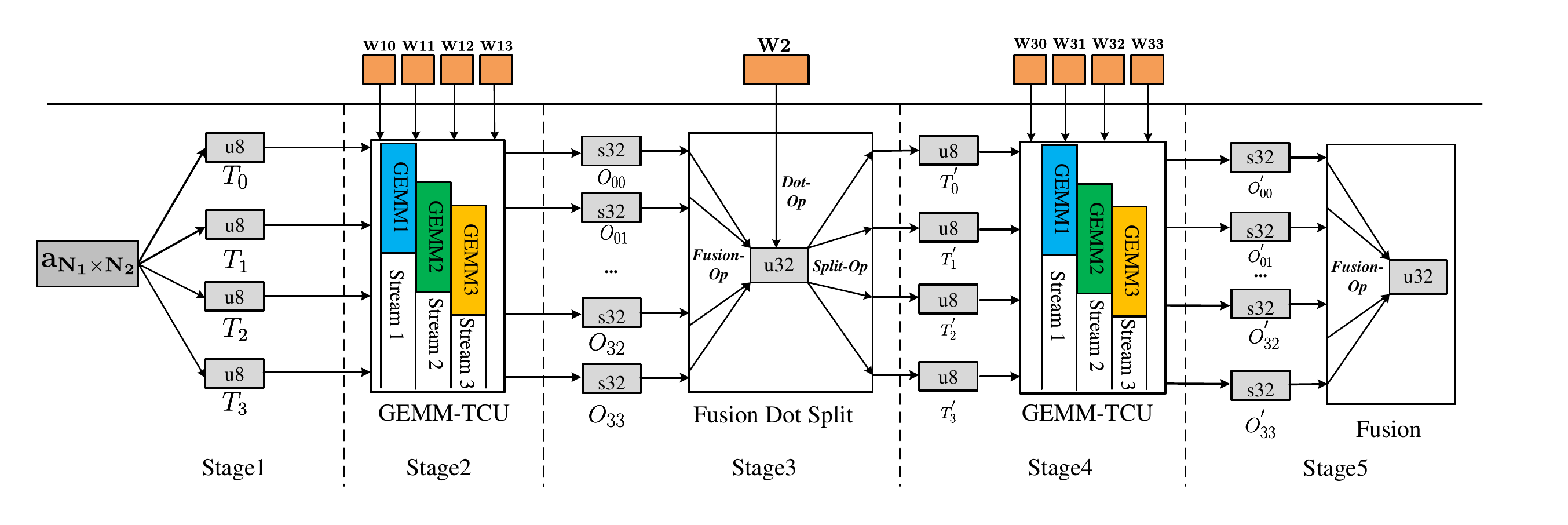}
\vspace{-25pt}
 \caption{Overall workflow of the NTT based on GEMM-TCU.}\label{fig:ntttcu}
\end{minipage}\hspace{-10pt}
\begin{minipage}[l]{0.34\linewidth}
    \includegraphics[width=\linewidth]{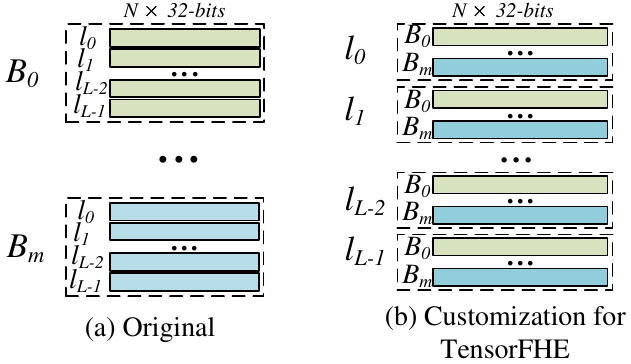}
    \vspace{-20pt}
    \caption{Data layout optimization in {\SolutionName}.}\label{fig:dataorg}
    \end{minipage}
    \vspace{-15pt}
\end{figure*}


Since the \emph{twiddle factor matrices} can be reused in all NTT kernels, their segmentation can be fulfilled as the pre-processing of the TCUs-based NTT kernels. The workflow of the TCUs-based NTT kernels is as shown in Figure \ref{fig:ntttcu}; it is composed of the following sequential stages:

\subsubsection{Stage1-Input Matrix Segmentation}

In this stage, we segment the input $a_{N_1\times N_2}$ (mentioned in Eq. \ref{eq:ntt_gemm}) into four small-scale matrices (denoted as $T_0$ to $T_3$) according to the process as shown in Figure \ref{fig:U32ToU8}. This stage is fulfilled with parallel threads on the CUDA cores; each thread fetches four consecutive 32-bit elements with the aim of maximizing global memory bandwidth utilization. Note that, to improve the efficiency of the 
subsequential GEMMs, all output matrices ($T_0$ to $T_3$) are stored in a column-major.

\subsubsection{Stage2-TCU GEMM for $a_{N_1\times N_2}\times W_1$}

Based on $T_0$ to $T_3$, we fulfill the $a_{N_1\times N_2}\times W_1$ on the TCUs. In this stage, the workload distributed to the TCUs can be formulated as
\begin{equation}
    O_{ij}=W_{1i}\times T_j,~~~i,j\in [0,3]
\end{equation}
where $W_{1i}$ indicates the segmentation matrices of $W_1$. Overall, there are 16 matrix multiplications to be fulfilled. To maximize computation efficiency, we leverage the workload concurrency by assigning each GEMM to a separate stream~\cite{guo2020accelerating}. Besides, we also use the open-source CUTLASS library \cite{cutlass} to implement all matrix multiplications. Note that, since the datatype of TCU output is restricted as \emph{s32} and the higher 16-bits are filled with zeros, it is risk-free for overflow in this stage. Besides, the output matrices $O_{ij}$ are stored in row-major.

\subsubsection{Stage3-Fusion and GEMM with $W_2$}

The output matrices of \texttt{Stage2} ($O_{ij}$) are firstly fused as one by using the \emph{Booth multiplication algorithm}\cite{madrid1993modified}, which treats the corresponding elements from every matrix as partial accumulations. The elements of the fused matrix are in the type of u32. After the fusion, we conduct a \emph{Hadamard multiplication} for the fused matrix and the $W_2$. Then, we segment the result matrix of the \emph{Hadamard multiplication} into four \emph{u8} matrices stored in column-major ($T'_0$ to $T'_3$) for the next stage.

\subsubsection{Stage4-TCU GEMM with $W_3$}

Similar to \texttt{Stage2}, we fulfill the $O'_{ij}=T'_j\times W_{3i}, i,j\in [0,3]$ on the TCUs. Each matrix multiplication is assigned to a separate stream to maximize the hardware efficiency.

\subsubsection{Stage5-Fusion and Output}

Similar to \texttt{Stage 3}, the output matrices of \texttt{Stage4} ($O'_ij$) are fused into one \emph{u32} matrix, which is stored in row-major, using the \emph{Booth multiplication algorithm}. Then, the fused matrix is moduloed element-wise by $q$ with output as the NTT result. Note that, for the \texttt{INTT} kernel, an extra modular multiplicative inverse of N under $q$ is multiplied element-wise with the result matrix before output.

Overall, \texttt{Stage1}, \texttt{Stage3} and \texttt{Stage5} are fulfilled on the CUDA cores with parallel threads, while the \texttt{Stage2} and \texttt{Stage4} are fulfilled on TCUs. In this way, we can significantly improve the performance of the NTT kernel by fully utilizing the emerging hardware resources in GPGPU.

\subsection{Operation-Level Batching}


As we discussed in Section \ref{sec:motiv:low-utilization}, the GPGPU 
is seriously underutilized when running CKKS operations. 
Therefore, we exploit to improve the overall performance of a real workload and hardware utilization by batching multiple FHE operations.

With operation-level batching, all running kernels process the data with the same $L$, since they can reuse the same \emph{twiddle factor matrix} for NTT. However, the improperly designed data layout may impair the effectiveness of the batching. As shown in Figure \ref{fig:dataorg}(a), $m$+$1$ CKKS operations are batched and their required data are originally stored in the manner of $(B,~L,~N)$. For each batching operation (i.e., $B_0$, ..., $B_m$), there are $L$ data entries stored in a contiguous address space (denoted as a group) and the size of each entry is $N\times 32$-bits. During the execution, the corresponding data entries with the same $L$ value from all groups are packed. Because the data entries are from the discontinuous memory space, performance of the batching execution suffers from low bandwidth utilization during the data packing.

When fully utilizing the data parallelism, we optimize the data layout for the operation-level batching. As shown in Figure \ref{fig:dataorg}(b), we reorganize the data layout in the manner of $(L,~B,~N)$, which stores all data entries with the same $L$ in continuous memory space. Thus, the batched operations can load the continuous data block (in size of $(m+1)\times N\times 32$-$bits$) for data packing, which helps to maximize bandwidth utilization. In this way, basic kernels can automatically generate a large amount of CTAs with the packed data and utilize as many GPGPU resources as possible.

\vspace{-5pt}
\subsection{{\SolutionName} Implementation}


By integrating the kernels and the above mentioned techniques, we can get an implementation of {\SolutionName}, which includes the following layers:
\begin{itemize}[wide, labelwidth=!, labelindent=0pt,noitemsep]
    \item \textbf{API Layer} runs on the CPU, which collects and decomposes the requests for FHE operations from the user applications. The decomposed requests are in the manner of workflow, which consists of the basic kernels. Then, the API layer automatically generates the best batch size for the different involved kernels according to the hardware resources. Finally, the API layer sequentially invokes the kernels in the workflow with proper batch size.
    \item \textbf{Kernel Layer} consists of various arithmetic kernels that run on the GPGPU. The kernel layer receives the invoking request from the API layer and returns the results of the requested operations. Note that the intermediate results of the kernels are stored in the GPGPU VRAM.
\end{itemize}

Based on such implementation, {\SolutionName} fully utilizes the hardware resources and efficiently supports all CKKS operations (including \texttt{Bootstrap}) 
on a single GPGPU.

\section{Methodology}\label{sec:methodology}
\begin{table}[t]
    \centering
    \caption{Platform Configurations.}\label{tab:platform_config}
    \vspace{-5pt}
    \begin{tabular}{|c|c|}\hline
        CPU &  Intel(R) Xeon(R) CPU E5-2678 v3 @ 2.50GHz \\ \hline
        GPGPU &  NVIDIA A100-SXM-40GB\\ \hline
        Memory & 512GB\\ \hline
        Misc. & CUDA 11.0; Pytorch 1.7.0; cuPy 9.6.0\\ \hline
    \end{tabular}
    \vspace{-10pt}
\end{table}

\begin{table}[t]
    \centering
    \caption{Compared Schemes.}\label{tab:designs}
    \vspace{-5pt}
    \begin{tabular}{|c|c|l|}\hline
        \textbf{Type} & \textbf{Design} & \textbf{Description} \\ \hline \hline
        CPU & Baseline\cite{craterlake} & AMD Ryzen 3975WX \\ \hline
       
        GPU & PrivFT\cite{privft} & NVIDIA Tesla V100, VRAM=16GB\\ \hline
     GPU & 100x\cite{100x} & NVIDIA Tesla V100, VRAM=16GB\\ \hline
        FPGA & HEAX\cite{heax} & Stratix 10 GX 2800, 11.7KB RFs\\ \hline
        \multirow{2}{*}{ASIC} & \multirow{2}{*}{F1+\cite{f1}} & 16 compute cluster (16 NTT FUs,\\ 
        && 32 Mul FUs, 16 Add FUs), 64MB RFs. \\  \hline
        \multirow{2}{*}{ASIC} & \multirow{2}{*}{CraterLake\cite{craterlake}} & 1$\times$ CRB FUs, 2$\times$NTT FUs,5$\times$ Mul \\ 
        &&FUs and 5$\times$ Add FUs, 256MB RFs\\ \hline
        \multirow{2}{*}{ASIC} & \multirow{2}{*}{BTS\cite{bts}} & 2048 $\times$ (NTT FUs,Add FUs\\ 
        &&Mul FUs), 512MB RFs\\ \hline
        \multirow{2}{*}{ASIC} & \multirow{2}{*}{ARK\cite{ark}} & \multirow{1}{*}{4$\times$ BConv FUs, 4$\times$NTT FUs, 4$\times$ }\\ 
        &&Auto FUs, 4$\times$ MAD FUs 512MB RFs\\ \hline \hline
        \multirow{1}{*}{GPGPU} & TensorFHE-NT & \RebuttalChange{Batching, NTT with \emph{butterfly operations}} \\ \cline{2-3}
        \multirow{2}{*}{(A100)}& \RebuttalChange{TensorFHE-CO} & \RebuttalChange{Batching, NTT with GEMMs}\\ \cline{2-3}
        & TensorFHE & \RebuttalChange{Batching, NTT with TCUs} \\ \hline
    \end{tabular}
    \vspace{-15pt}
\end{table}


\begin{table}[t]
    \centering
    \caption{CKKS Parameters used in the experiments.}\label{tab:default_parameter}
    \vspace{-5pt}
    \begin{tabular}{|c|c|c|c|c|c|}\hline
          & \textbf{N} & \textbf{L} & \textbf{K} & \textbf{logPQ} & \textbf{batch\_size}  \\ \hline \hline
        Default & $2^{16}$& 44&1 &1306 & 128\\ \hline
        ResNet-20 & $2^{16}$& 29& 1& 840& 64 \\ \hline
        Logistic Regression & $2^{16}$&38 & 1& 1092& 64\\ \hline
        LSTM & $2^{15}$& 25&1 &728 &32 \\ \hline
        Packed Bootstrapping & $2^{16}$& 57& 1& 1624&32 \\ \hline
    \end{tabular}
    \vspace{-15pt}
\end{table}

We implement our proposed {\SolutionName} based on CUDA \Tofill{11.0}\cite{cuda11} and PyTorch \Tofill{1.7}\cite{pytorch}. Then, we evaluate the {\SolutionName} on a high-end server equipped with one NVDIA A100 GPGPU. The detailed configurations of the platform are as shown in Table \ref{tab:platform_config}. Besides, we use \Tofill{NVIDIA Nsight}\cite{nsight} to monitor the SM occupancy during the execution. 

To fully evaluate our proposed {\SolutionName}, we compare it with a series of previous works based on various devices. The involved designs are as shown in Table \ref{tab:designs}. To compare {\SolutionName} with works based on CPU, GPGPU and FPGA, we measure the performance of the key kernel and the operations, including \texttt{NTT, HADD, HMULT, HROTATE} and \texttt{Bootstrap}. For the comparison with ASIC accelerators, we evaluate the {\SolutionName} by using four CKKS programs with state-of-the-art implementations and measure the overall execution time. Note that we directly collect data from the literature for the previous works, and use a dash ‘-’ to represent data not mentioned in the literature. 

The programs used for the comparison with ASIC accelerators are as follows:
\begin{itemize}[wide, labelwidth=!, labelindent=0pt,noitemsep]
    \item \textbf{ResNet-20\cite{resnet20}}. We implemented this DNN model for image recognition with FHE. In our experiment,  64 encrypted images are packed as the input. 
    \item   \textbf{Logistic Regression\cite{helr}}. We test the HELR algorithm with 16384 samples, and batch-encode 128 samples in one polynomial. We execute it for 14 iterations, which three bootstrapping operations are required.
     \item   \textbf{LSTM\cite{lstm}}. This is an NLP model for analyzing the contextual information of a sentence. We implement this model in FHE as \cite{lstm}, which includes 128 cells, and the dimension for each word embedding used in this model is 128. We pack 32 sentences in parallel as the input.
     \item \textbf{Packed Booststrapping\cite{mouchet2020lattigo}}. This example performs packed booststrapping operations with the same configuration as \cite{craterlake}. The 
     ciphertext with $N=64k$ restores its security level to L=57. In our experiment, we perform 32 ciphertexts in parallel for the bootstrapping operations. 
\end{itemize}

Table \ref{tab:default_parameter} presents the CKKS parameters used in the experiments. Without specific description, we use parameters in \texttt{Default} for all experiments.  

\section{Results}\label{sec:result}






In order to evaluate the effectiveness of {\SolutionName}, we provide a series of comparisons between {\SolutionName} and various previous works. For the comparison with the works on CPU, GPGPU and FPGA, we focus on the performance of FHE operations and the throughput of the critical kernels. On the other hand, we measure the performance of the real workload for comparison with the ASIC accelerators. We also provide a sensitivity study of the different FHE parameters.

\subsection{\RebuttalChange{The effectiveness of NTT optimization}}

\begin{figure}[t]
    \centering
    \includegraphics[width=0.85\linewidth]{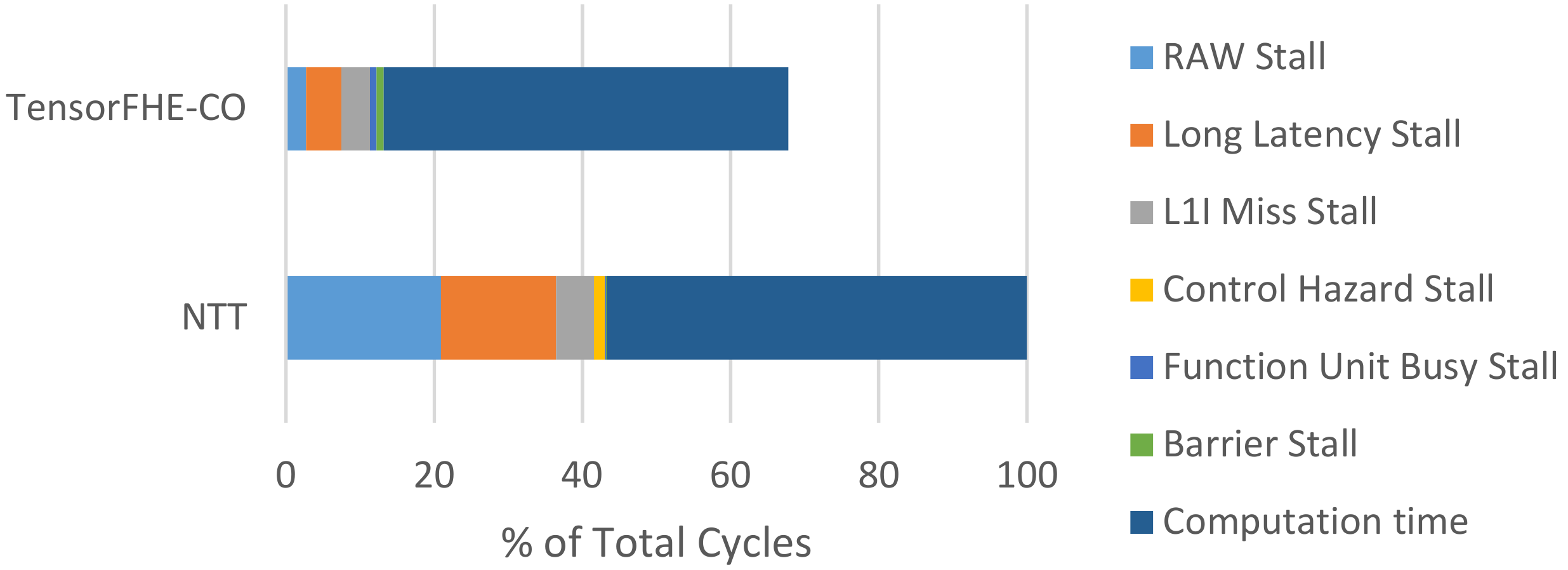}
    \vspace{-10pt}
    \caption{\RebuttalChange{Normalized pipeline execution time breakdown comparison of the NTT implementations.}}
    \label{fig:results:raw-stall}
    \vspace{-15pt}
\end{figure}

\RebuttalChange{To evaluate the effectiveness of NTT optimization in {\SolutionName}, we run the optimized NTT kernel (used in the TensorFHE-CO) on the simulation platform as introduced in Section \ref{sec:motiv:ntt}, and then compare the execution time breakdown to the state-of-the-art implementation\cite{dai2015cuhe}. As shown in Figure \ref{fig:results:raw-stall}, {\SolutionName} significantly reduces the \emph{RAW stall} by \Tofill{18.1\%} and the \emph{Long Latency Stall} by \Tofill{10.8\%}. Note that, although transformation from \emph{butterfly operations} to the matrix operations increases the computation time by \Tofill{1.2\%}, 
the overall performance of NTT is still improved by \Tofill{32.3\%} due to the reduction of the various pipeline stalls. Besides, we also validate the correctness of the optimized NTT kernel by using successive NTT and INTT kernel. We compare the NTT input with the INTT ouput, and find that the two values are exactly the same, which proves the correctness of our optimization.}

\vspace{-5pt}
\subsection{{\SolutionName} vs. CPU, GPGPU and FPGA}

\subsubsection{Performance}

As shown in Table \ref{tab:performance_operation}, {\SolutionName} achieves significant speedups over the implementation based on CPU, which is $397.1\times$ faster with \texttt{HMULT} and up to $1035.8\times$ faster with \texttt{HADD}. Moreover, for the \texttt{HMULT} and the \texttt{HROTATE}, which involve expensive NTT computations, {\SolutionName}-NT achieves the speedup over 100x\cite{100x} (the state-of-the-art CKKS implementation on GPGPU) for \Tofill{1.04$\times$} and \Tofill{1.02$\times$}, respectively. Such improvement proves that hardware efficiency significantly impacts FHE performance, \RebuttalChange{which is mainly caused by the batching execution. Besides, {\SolutionName}-CO helps to achieve \Tofill{$1.35\times$} and \Tofill{$1.41\times$} speedup over 100x\cite{100x} for the \texttt{HMULT} and the \texttt{HROTATE} operation, which is directly improved by reduction of the RAW stalls and the modulo operations. Besides, {\SolutionName} performs \Tofill{2.49$\times$} and \Tofill{$1.84\times$} speedup over {\SolutionName}-NT and {\SolutionName}-CO, which shows the impact of emerging TCUs on NTT operations.} \RebuttalChange{Moreover, to directly compare the performance, we re-run the {\SolutionName} on NVIDIA V100 (same as 100x\cite{100x}), and observe that the {\SolutionName} still performs faster than 100x\cite{100x}, which is up to \Tofill{$5.26\times$} speedup for the \texttt{RESCALE}.} Furthermore, as shown in Table \ref{tab:performance_boot}, {\SolutionName} achieves $1.3\times$ speedup over 100x for \texttt{Bootstrap}, which is the key to fulfilling complex FHE workloads.

We also compare {\SolutionName} to the state-of-the-art implementation of FHE on FPGA. As shown in Table \ref{tab:performance_ckks_with_heax}, TensorFHE achieves 4.9$\times$ speedup on average for the (i)NTT kernel. For the \texttt{HMULT} operation, TensorFHE can also achieve 1.46$\times$ speedup for Set\_C. However, with the small polynomial length of Set\_A, TensorFHE is slower than HEAX by about $10\%$, which implies that algorithm complexity reduction provides more advantages than massive parallelism for workloads requiring less computation.

\begin{table}[t]
    \centering
    \footnotesize
    \caption{Operation Delay Comparison of {\SolutionName}, CPU and GPGPU on micro FHE benchmarks (ms)
    }\label{tab:performance_operation}
    \vspace{-5pt}
    \begin{tabular}{|c|c|c|c|c|c|}\hline
 &\texttt{HMULT} 	&        \texttt{HROTATE} 	 &        \texttt{RESCALE} 	&         \texttt{HADD} 	&         \texttt{CMULT} \\ \hline
 CPU\cite{100x} &	338s&	330s&	18611	&3609	&3356\\ \hline
 

PrivFT 	\cite{privft}&7153&	-&	208	&24	&21\\ \hline
 100x \cite{100x}&	2227&	2154	&81&	26&	22\\ \hline
\hline
 \textbf{{\SolutionName}-NT} &	\textbf{2124}	& \textbf{2111} &	\textbf{35}&	\textbf{6}	&\textbf{7.7}\\ \hline
  \textbf{\RebuttalChange{{\SolutionName}-CO}} &	\RebuttalChange{\textbf{1651.2}}	& \RebuttalChange{\textbf{1523.2}} &	\RebuttalChange{\textbf{9.2}}&	\RebuttalChange{\textbf{6}}	&\RebuttalChange{\textbf{7.7}}\\ \hline
   \textbf{\RebuttalChange{{\SolutionName}(V100)}} &  \RebuttalChange{\textbf{1296.6}} &	\RebuttalChange{\textbf{1254.4}} &	 \RebuttalChange{\textbf{15.4}} &	 \RebuttalChange{\textbf{10.2}} &	 \RebuttalChange{\textbf{11.5}} \\ \hline
 \textbf{{\SolutionName}(A100)} &  \textbf{851} &	\textbf{852} &	 \textbf{7.7} &	 \textbf{6} &	 \textbf{7.7} \\ \hline

    \end{tabular}
    \vspace{-10pt}
\end{table}

\begin{table}[t]
    \centering
    \caption{\RebuttalChange{Execution Time Comparison for the \texttt{Bootstrap} (in seconds) \\ (N=$2^{16}$, L=34, dnum=5 and batch\_size = 128).}}\label{tab:performance_boot}
    \vspace{-5pt}
    \begin{tabular}{|c|c|c|c|c|c|}\hline
     \multirow{1}{*}{CPU} &\multirow{1}{*}{ GPGPU }&\multirow{1}{*}{100$\times$}& \multirow{1}{*}{Tensor }&\RebuttalChange{\multirow{1}{*}{Tensor}}&\multirow{1}{*}{Tensor} \\ 
     \cite{100x}&baseline\cite{100x}&\cite{100x}&FHE-NT&\RebuttalChange{FHE-CO}& FHE\\ \hline
     \multirow{2}{*}{10168}&	\multirow{2}{*}{54904}&\multirow{2}{*}{42016}&\multirow{2}{*}{76731}&\RebuttalChange{\multirow{2}{*}{70762}}&\multirow{2}{*}{32058}\\ 
    &&&&& \\ \hline
    \end{tabular}
    \vspace{-15pt}
\end{table}

\begin{table}[t]
    \centering
    \caption{Performance  Comparison for {\SolutionName} and HEAX. Taking the throughput of the key kernels and operation as metric.
    }
    \vspace{-5pt}
    \label{tab:performance_ckks_with_heax}
    \begin{tabular}{|c|c|c|c|c|}\hline
    
&& Set\_A 	&    Set\_B 	&     Set\_C \\ \hline
\hline
\multirow{3}{*}{\#NTT/second}&CPU\cite{heax}& 7222& 	3437& 	1631\\ \cline{2-5}
&HEAX\cite{heax}&  	195313& 	90144& 	41853\\ \cline{2-5}
&TensorFHE&  	910134& 	449974& 	209337\\ \hline 
\multirow{3}{*}{\#INTT/second}&CPU\cite{heax}& 7568& 	3539& 	1659\\ \cline{2-5}
&HEAX\cite{heax}& 	195313& 	90144& 	41853\\ \cline{2-5}
&TensorFHE& 913267& 	449084& 	209178\\ \hline
\multirow{3}{*}{\#\texttt{HMULT}/second}&CPU\cite{heax}& 420& 	84& 	15\\ \cline{2-5}
&HEAX\cite{heax}& 	97656& 	22536& 	2616\\ \cline{2-5}
&TensorFHE& 88048& 	 27564& 	 3825\\   \hline




    \end{tabular}
    
\begin{tablenotes}
\footnotesize
\item{Set\_A}: N=$2^{12}$, $log_{pq}$ = 108, K=2; {Set\_B}: N=$2^{13}$, $log_{pq}$ = 217, K=4; {Set\_C}: N=$2^{14}$, $log_{pq}$ = 437, K=8.
\end{tablenotes}
\vspace{-10pt}
\end{table}


\subsubsection{Execution Time Breakdown}

\begin{figure}[t]
    \centering
    \includegraphics[width=0.9\linewidth]{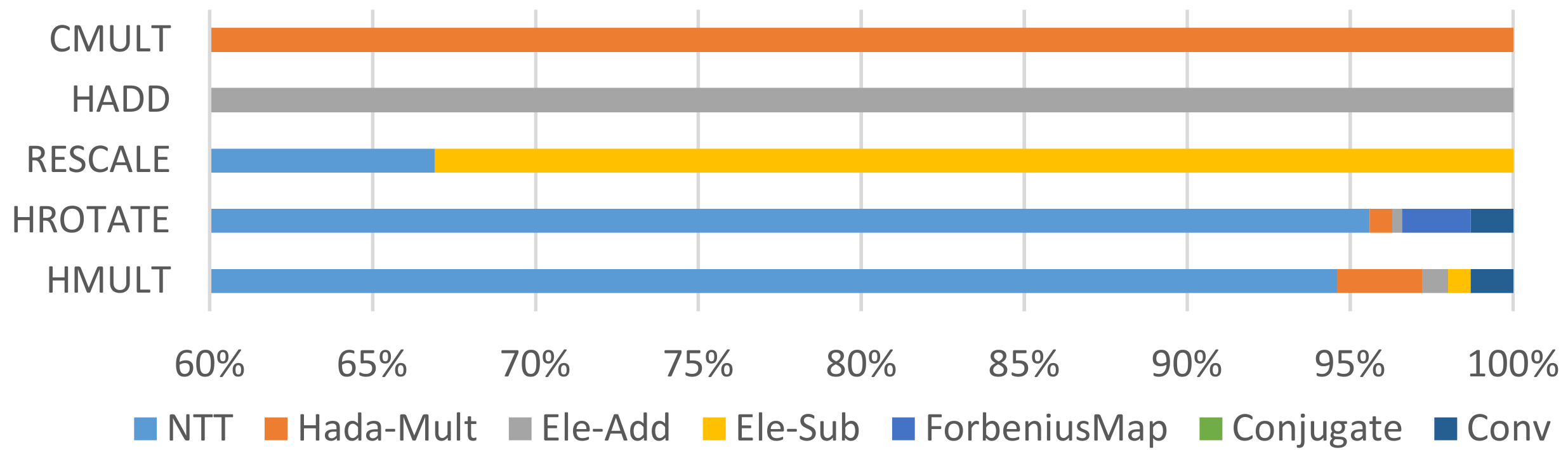}
    \vspace{-10pt}
    \caption{\RebuttalChange{Execution Time Breakdown for the FHE Operations.}}\label{tab:operation_breakdown}
    \vspace{-15pt}
\end{figure}

Table \ref{tab:operation_breakdown} presents the execution time breakdown of different {\SolutionName} operations. Not surprisingly, the NTT kernels occupy the most significant proportion in \texttt{HMULT} and \texttt{HROTATE}, that is, $92.1$\% and $95.4\%$, respectively. This indicates that, though current {\SolutionName} provides remarkable performance improvement for NTT/INTT kernels, there are still more opportunities for further improvement. On the other hand, the non-NTT kernels only take a small part of the time, which indicates that the fully parallel scheme using {\SolutionName} has already achieved good performance for the non-NTT kernels. 


\subsubsection{Hardware Utilization}

\begin{table}[t]
    \centering
    \caption{GPGPU Occupancy of the {\SolutionName} Operations. 
    }\label{tab:operation_utlization}
    \vspace{-5pt}
    \begin{tabular}{|c|c|c|c|c|c|}\hline
    &\texttt{HMULT}&\texttt{HROATE}&\texttt{RESCALE}&\texttt{HADD}&\texttt{CMULT} \\ \hline
     Occupancy &90.3\%&	90.1\%&	88.9\%	&85.3\%&	88.1\% \\ \hline
    \end{tabular}
    \vspace{-10pt}
\end{table}

As we discussed in Section \ref{sec:motiv}, FHE performance suffers from low GPGPU occupancy. As shown in Table \ref{tab:operation_utlization}, {\SolutionName} achieves remarkable improvement for hardware occupancy by using the batching techniques. Especially, for the \texttt{HMULT} and the \texttt{HROTATE}, the GPGPU occupancy is over $90\%$. Note that for the GPGPU occupancy calculation, we consider the TCUs for the \texttt{HMULT} and the \texttt{HROTATE}, while only considering the CUDA cores for the other operations.


\subsection{\RebuttalChange{Evaluation with Real Workload}}

\subsubsection{Performance}

As shown in Table \ref{tab:performance_app}, for some specific applications, {\SolutionName} provides comparable performance with the state-of-the-art FHE accelerators based on ASIC. Especially, for the \texttt{LR}, {\SolutionName} achieves $1625.6\times$ and $2.9\times$ speedup over the CPU and F1+, respectively. When compared with the other ASIC accelerators (i.e., BTS, ARK and CraterLake), there is still a performance gap, up to $40\times$. However, since {\SolutionName} is a high-performance FHE implementation based on GPGPU, there are still many chances to further improve the performance with an acceptable overhead, such as shifting to the platforms with multiple GPGPUs. 



\begin{table}[t]
    \centering
    \caption{
    Performance Comparison for the Full FHE Workloads. Taking the execution time (in seconds) as the metric.}\label{tab:performance_app}
    \vspace{-5pt}
    \begin{tabular}{|c|c|c|c|c|}\hline
&\multirow{2}{*}{\texttt{ResNet-20}}&\multirow{2}{*}{\texttt{LR}}&\multirow{2}{*}{\texttt{LSTM}}&\texttt{Packed} \\
    &&&&\texttt{Bootstrapping} \\ \hline
 CPU\cite{craterlake} &	 88320	&	 22784	&	 27488	&	 550.4  \\ \hline
 F1+\cite{f1} 	& 172.3	&	40.9	&	82.3&	 1.8\\ \hline
 CraterLake \cite{craterlake}	& 15.9	&	7.6	&	4.4&	0.1\\ \hline
 BTS \cite{bts}&	122.2	&	1.8&	 -&	 -\\ \hline
 ARK \cite{ark}&	18.8	&	0.49&	 -&	 -\\ \hline
 \RebuttalChange{ \texttt{100x}$^{*}$ \cite{100x}}&	 \RebuttalChange{602.9}	&	 \RebuttalChange{49.6}&	 -&	  \RebuttalChange{36.9}\\ \hline
 {\SolutionName} &	316.1&	14.1	&123.1&	13.5 \\ \hline

    \end{tabular}
    \begin{tablenotes}[flushleft]\footnotesize{
    \item \RebuttalChange{*The execution time of ResNet-20 and Packed-Bootstrapping of \texttt{100x} are estimated based on the executed numbers of HE operands.}
    }
    \end{tablenotes}
    \vspace{-15pt}
\end{table}

\subsubsection{Execution Time Breakdown}

\begin{figure}[t]
    \centering
    \includegraphics[width=0.9\linewidth]{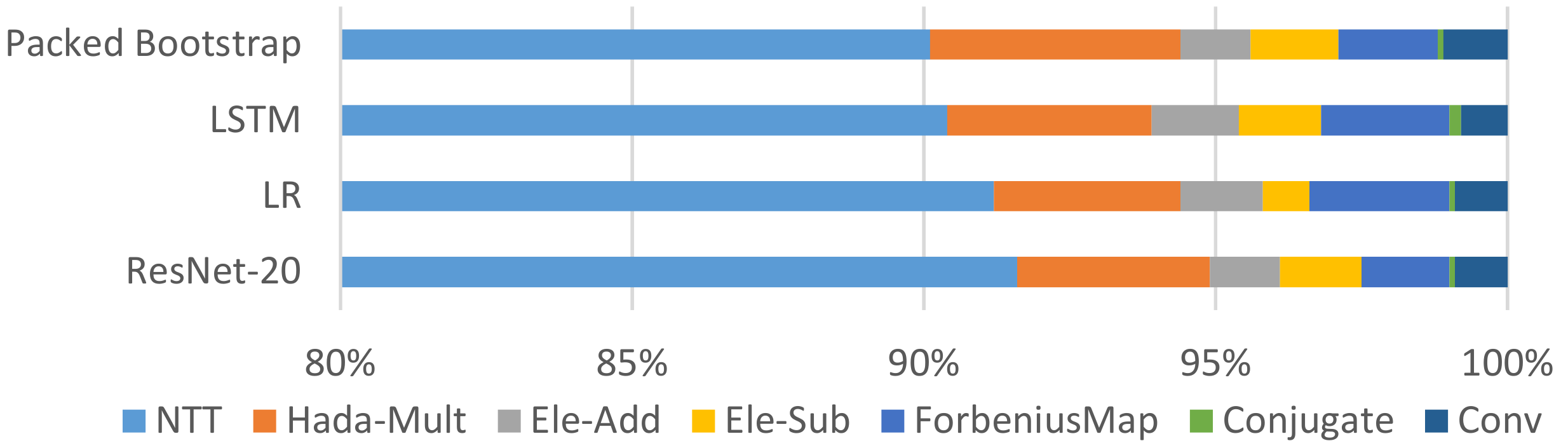}\vspace{-5pt}
    \caption{\RebuttalChange{Kernel-level Execution Time Breakdown
    }}\label{tab:appbreakdown_kernel}
    \vspace{-10pt}
\end{figure}

\begin{figure}[t]
    \centering
    \includegraphics[width=0.9\linewidth]{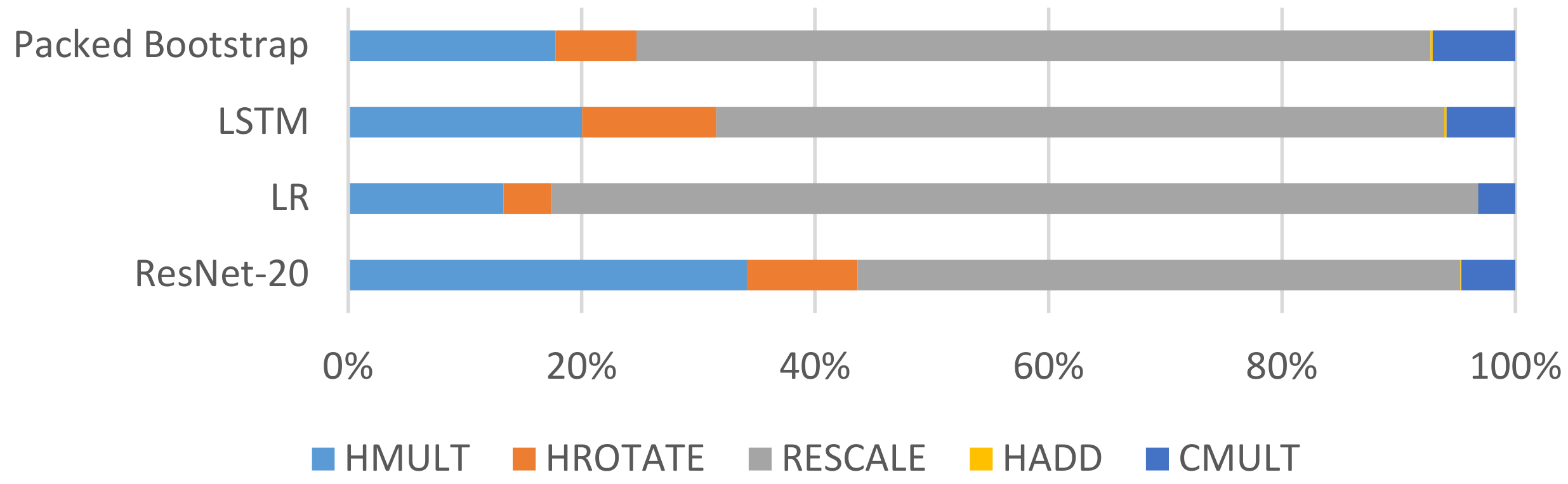}
    \vspace{-10pt}
    \caption{\RebuttalChange{Operation-level Execution Time Breakdown 
    }}\label{tab:appbreakdown_operation}\vspace{-15pt}
\end{figure}

We calculate the kernel-level and operation-level execution time breakdown for executing the real workload. The results are as shown in Table \ref{tab:appbreakdown_kernel} and Table \ref{tab:appbreakdown_operation}, respectively. As shown in Figure \ref{tab:appbreakdown_kernel}, the NTT kernels take the largest proportion of the execution time for all workloads, which is up to $92.8\%$ in \texttt{LR}. Considering the granularity of the operations, \texttt{HROTATE} becomes the most time-consuming operation, which is frequently used and contains a mass of \texttt{NTT} kernels.    

\vspace{-5pt}

\subsection{\RebuttalChange{Energy}}\vspace{-5pt}

\begin{table}[t]
    \centering
    \caption{\RebuttalChange{Energy Efficiency of {\SolutionName}}
    }\label{tab:energy}
    \vspace{-5pt}
    \begin{tabular}{|c|c|c|c|c|}\hline
    \multicolumn{5}{|c|}{Energy Efficiency of CKKS operations~(OPs/W)} \\ \hline
    \texttt{HMULT}&\texttt{HROATE}&\texttt{RESCALE}&\texttt{HADD}&\texttt{CMULT} \\ \hline
     0.57&	0.57&	66.67	&81.30&	66.67 \\ \hline \hline
     \multicolumn{5}{|c|}{Energy Consumption of full workload~(J/iteration)} \\ \hline
     &\multirow{1}{*}{\texttt{ResNet}}&\multirow{2}{*}{\texttt{LR}}&\multirow{2}{*}{\texttt{LSTM}}& \multicolumn{1}{c|}{\texttt{Packed }}\\ 
     &\texttt{-20}&&&\texttt{Bootstrap} \\ \hline
     \textbf{ARK\cite{ark}}& 32.5&	19.8&	-	&	\multicolumn{1}{c|}{-} \\ \hline
     \textbf{CraterLake\cite{craterlake}}& 79.7&	38.1&	44.2	&	\multicolumn{1}{c|}{1.3} \\ \hline
    \textbf{TensorFHE}& 1320&	58.27&	1015.3	&	\multicolumn{1}{c|}{111.3} \\ \hline
    \end{tabular}
    \vspace{-5pt}
\end{table}

\RebuttalChange{To evaluate the energy efficiency of {\SolutionName}, we use \Tofill{NVIDIA-SMI} to monitor the run-time GPGPU power, which is stable at \Tofill{264} watts in our experiments due to the high hardware utilization. The energy efficiency of different CKKS operations is as shown in Table \ref{tab:energy}, which is up to $81.30$ OPs/W in \texttt{HADD}. Besides, we also report the energy efficiency of various workload, which varies from 111.3 J/iteration (\texttt{Packed Bootstrap}) to 1320 J/iteration (\texttt{ResNet-20}). Table \ref{tab:energy} also shows that the energy consumption of TensorFHE is higher than the ASICs, due to the higher power of GPGPU.}

\vspace{-5pt}
\subsection{Sensitivity Study to the Batch Size}


\begin{figure}[t]
    \centering
    \begin{minipage}[l]{\linewidth}
\includegraphics[width=0.9\linewidth]{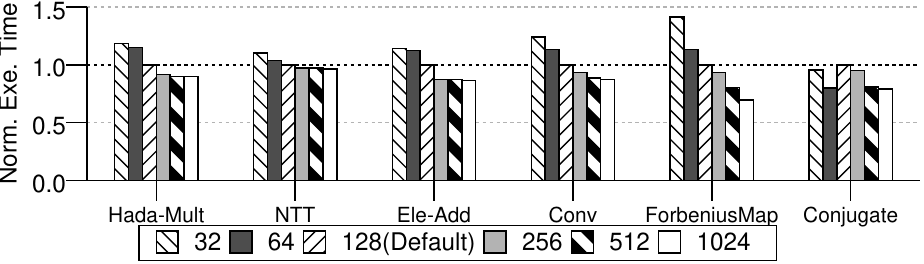}
    \vspace{-10pt}
    \caption{Impact of batch size on execution time.}\label{fig:sent_batch_size}
    \vspace{10pt}
\end{minipage}

\begin{minipage}[l]{\linewidth}
    \includegraphics[width=0.9\linewidth]{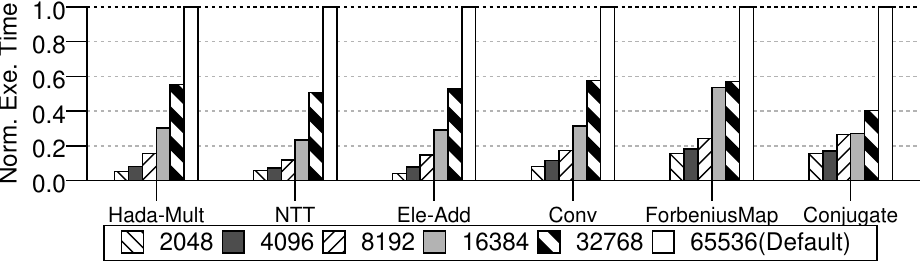}
    \vspace{-10pt}
    \caption{Sensitivity to \emph{polynomial length} of {\SolutionName}.}\label{fig:sent_N}
    \end{minipage}
    \vspace{-15pt}
\end{figure}


\emph{Batch Size} refers to the number of identical FHE operations simultaneously executed. As discussed in Section \ref{sec:design}, the twiddle factor matrix can be shared by all NTT kernels with the same $N$ and $q$. Therefore, a larger batch size allows more operations to fully utilize the reusable data and the hardware resources, which helps to improve the performance of the real workload based on FHE schemes. However, as the batched operations increase, the requirement for the VRAM capacity also grows, which is caused by the increasing intermediate data. Therefore, the batch size of {\SolutionName} is mainly determined by the VRAM capacity of the GPGPU.

Figure \ref{fig:sent_batch_size} presents the impact of the batch size on the overall performance of {\SolutionName}. Here we use the \texttt{Default} configuration as shown in Table \ref{tab:default_parameter}, and only change the value of the batch size (denoted as $BS$). As the $BS$ increases from 32 to 1024, the performance of all kernels is improved. For example, the \texttt{ForbeniusMap} achieves the best performance improvement with \texttt{BS=1024}, which achieves \Tofill{31.4\%} performance improvement compared with \texttt{BS=128}.
However, Figure \ref{fig:sent_batch_size} also shows that different kernels have various optimal $BS$ values. To balance the workload of different kernels, in this paper, we take $BS=128$ as the default configuration. 



\subsection{Sensitivity Study to the Polynomial Length}



The polynomial length (denoted as $N$) of {\SolutionName} refers to the number of coefficients in the polynomials used to represent the input data. A larger $N$ means that the input data will be represented as a longer polynomial with a higher security level guarantee\cite{han2020better}. However, it also involves a higher computational overhead. Here we fix other attributes and vary the value of $N$ from $65536$ to $2048$. Figure \ref{fig:sent_N} shows the performance, measured as the execution time.

We can see that {\SolutionName} with $N=65536$ performs considerably worse than other configurations. This is because a longer polynomial increases the computational workloads for all kernels, especially for the NTT kernels, which also require a much larger twiddle factor matrix. On the other hand, {\SolutionName} runs much faster with smaller $N$. Especially, as the $N$ decreases from 65536 to 2048, the NTT kernel gains \Tofill{20.6$\times$} speedup, since the computational workload is reduced by \Tofill{97\%}. However, this also decreases the security level. In this paper, we use a default configuration with $n=65536$ to provide the acceptable security level for the real workloads.

\section{Discussion}\label{sec:discussion}


\textbf{Further Performance Improvement.} Although the current {\SolutionName} achieves significant performance improvement, there are still several opportunities for further optimization, which can be mainly summarized in two aspects: on the one hand, the arithmetic kernels can be further optimized by improving the parallelism or using a streaming pipeline; on the other hand, extending {\SolutionName} to the platform with multiple GPGPUs would help to increase the batch size, which improves the performance of complex workloads by further improving the throughput of CKKS operations. We will implement these characteristics in future {\SolutionName}.

\textbf{Generality.} In this paper, we focus on the emerging CKKS scheme. However, our proposed {\SolutionName} can also support other FHE schemes, such as BFV\cite{bfv} and BGV\cite{BGV}. For a new FHE scheme, the \emph{API Layer} provides new APIs to the user applications and invokes the required kernels, while the \emph{Kernel Layer} would require new arithmetic kernels to support different algorithms.

\RebuttalChange{\textbf{Security Vulnerability.} {\SolutionName} achieves significant performance improvement on GPGPU without changing the key algorithm of CKKS. Therefore, though the GPGPU might be untrusted, security can still be guaranteed since all the computations are fulfilled on the encrypted data.}

\section{Related Work}\label{sec:related}
\textbf{GPU-based FHE acceleration.} Several previous works have been proposed to accelerate various FHE schemes on GPGPU\cite{100x,privft,fvgpu,fhecrt,gpubgv}. \RebuttalChange{Privft\cite{privft} the first CKKS implementation with the RNS-variant on GPGPU. It directly maps the original CKKS algorithm to the GPGPU using the data tiling. However, it failed to support bootstrap and provided no deep insights. Based on this work, Ref \cite{fhecrt} uses the Barret Reduction method\cite{barret} and CRT to improve the performance of FHE schemes on GPGPU. 100x\cite{100x} is the first CKKS implementation on GPGPU with bootstrap support. It analyzes the bottleneck of memory accesses and then improves the CKKS performance using the optimized function kernel with fewer memory accesses. Ref \cite{fvgpu} accelerates the FV scheme on GPGPU, which uses the CRT and discrete Galois transformation\cite{al2018efficient} to avoid the additional multi-precision arithmetic operations. Ref \cite{gpubgv} improves the performance of the BGV scheme on GPGPU, which optimizes the modular operations with the Barret Reduction method\cite{barret} and overlaps the data transfer between the host and GPU with the computation. These previous works focus on the optimization of memory accessing, while our TensorFHE makes an effort to improve the performance of CKKS by enhancing hardware efficiency and utilization with algorithm optimizations.}

\textbf{Hardware-based FHE accelerator.} Previous works also explore to accelerate FHE schemes by using the FPGA\cite{heax,cousins2016designing,turan2020heaws,cousins2014fpga} or ASIC accelerators\cite{bts,craterlake,ark,f1}. Due to the limited on-chip hardware resources, the FPGA-based approaches focus on the acceleration for the specific key operations, such as the NTT/INTT\cite{heax,cao2014high}, the ModMult\cite{heax,cao2013accelerating}, and the ModAdd\cite{cao2014high}. However, for the real workloads, the overall performance suffers from the frequently data transfer caused by the partial acceleration support. On the other hand, ASIC accelerators support all FHE operations and significantly improve the performance of FHE-based workloads\cite{bts}. But, all of these works require the huge on-chip storage (i.e., 256MB for F1+\cite{f1} and CraterLake\cite{craterlake}, 512MB for BTS\cite{bts} and ARK\cite{ark}), which leads to the over-high implementation cost. Our proposed {\SolutionName} proves that, with algorithm optimization based on the understanding of micro-architectural characteristics, it is possible to achieve comparable performance for the FHE workloads on GPGPUs.  

\section{Conclusion}\label{sec:conclusion}
In this paper, we provide a detailed analysis of the inefficiency of running FHE operations on GPGPU and then propose {\SolutionName}, which is a pure software FHE acceleration solution based on single GPGPU. {\SolutionName} uses a hierarchical model to reconstruct the CKKS, which decomposes the FHE operations into a series of reusable kernels.
Then, {\SolutionName} optimizes the algorithm of the NTT kernel to fit it with the emerging TCUs and utilizes the regular CUDA cores to accelerate the other kernels. Moreover, to fully utilize the potential data parallelism of GPGPU, {\SolutionName} uses an operation-level batching technique to allow more FHE operations to be executed simultaneously. 
In this way, the proposed {\SolutionName} significantly improves the performance of FHE applications by executing more FHE operations in the certain period of time. The experiment results show that {\SolutionName} achieves much higher performance than all state-of-the-art FHE acceleration solutions on CPU, GPU and FPGA. More surprisingly, {\SolutionName} provides comparable performance when compared with the state-of-the-art ASIC FHE accelerators, and even achieves higher performance in some applications. Considering the high expense of ASIC implementation, we believe that {\SolutionName} can be a competitive candidate for applying FHE to the cloud computing servers in the near future, and will possibly inspire more works. 


\bibliographystyle{IEEEtranS}
\bibliography{refs}  

\end{document}